\documentclass[ALICE,manyauthors]{cernphprep}
\usepackage[english]{babel}
\usepackage{graphicx}
\usepackage{verbatim}
\usepackage{amsfonts}
\usepackage{makeidx}
\usepackage{color}
\usepackage{amsmath}
\usepackage{lineno}
\usepackage{epsfig}
\usepackage{epstopdf}
\usepackage{hyperref}
\usepackage{cite}
\usepackage{multirow}
\usepackage{subfigure}

\usepackage[comma,square,numbers,sort&compress]{natbib}

\newcommand{\sqrts}{\sqrt{s}}
\newcommand{\sqrtsNN}{\sqrt{s_{\scriptscriptstyle \rm NN}}}

\newcommand{\av}[1]{\left\langle #1 \right\rangle}

\newcommand{\MeV}{\mathrm{MeV}}
\newcommand{\GeV}{\mathrm{GeV}}
\newcommand{\TeV}{\mathrm{TeV}}

\newcommand{\gev}{\mathrm{GeV}}
\newcommand{\gevc}{\mathrm{GeV}/c}
\newcommand{\tev}{\mathrm{TeV}}

\newcommand{\mum}{\mathrm{\mu m}}

\newcommand{\mub}{\mathrm{\mu b}}

\newcommand{\PbPb}{\mbox{Pb--Pb}}

\newcommand{\pt}{p_{\rm T}}

\newcommand{\DtoKpi}{{\rm D}^0 \to {\rm K}^-\pi^+}
\newcommand{\DtoKpipi}{{\rm D}^+\to {\rm K}^-\pi^+\pi^+}
\newcommand{\DstartoDpi}{{\rm D}^{*+} \to {\rm D}^0 \pi^+}

\newcommand{\Dzero}{{\rm D^0}}
\newcommand{\Ds}{{\rm D^+_s}}

\newcommand{\Dstar}{{\rm D^{*+}}}

\newcommand{\Dplus}{{\rm D^+}}

\renewcommand{\d}{\mathrm{d}}

\renewcommand{\PbPb}{\mbox{Pb--Pb}}

\newcommand{\Raa}{R_{\rm AA}}

\newcommand{\RAA}{R_{\rm AA}}
\newcommand{\TAA}{T_{\rm AA}}

\begin{document}

\begin{titlepage}

\PHyear{2018}
\PHnumber{066}                 
\PHdate{6 April}              

\title{Measurement of $\Dzero$, $\Dplus$, $\Dstar$ and $\Ds$ production in $\PbPb$ collisions at $\mathbf{\sqrt{{\textit s}_{\rm NN}}~=~5.02~TeV}$}
\Collaboration{ALICE Collaboration\thanks{See Appendix~\ref{app:collab} for the list of collaboration members}}
\ShortAuthor{ALICE Collaboration} 
                     
\ShortTitle{D-meson production in Pb--Pb collisions at $\sqrtsNN=5.02~\tev$}

\begin{abstract} 
We report measurements of the production of prompt $\Dzero$, $\Dplus$, $\Dstar$ and $\Ds$ mesons in Pb--Pb collisions 
at the centre-of-mass energy per nucleon--nucleon pair $\sqrtsNN=5.02~\tev$, in the 
centrality classes 0--10\%, 30--50\% and 60--80\%. The D-meson production yields are measured at mid-rapidity ($|y|<0.5$)
as a function of transverse momentum ($\pt$). The $\pt$ intervals covered in central collisions are: $1<\pt<50~\gev/c$ for $\Dzero$, $2<\pt<50~\gev/c$ for $\Dplus$, 
$3<\pt<50~\gev/c$ for $\Dstar$, 
and $4<\pt<16~\gev/c$ for $\Ds$ mesons. 
The nuclear modification factors ($\RAA$)
for non-strange D mesons ($\Dzero$, $\Dplus$, $\Dstar$) show minimum values of about 0.2 for $\pt=6$--$10~\gev/c$ in the most central collisions and are compatible within uncertainties with those measured at $\sqrtsNN=2.76~\tev$. For $\Ds$ mesons, the values of $\RAA$ are larger than those of non-strange D mesons, but compatible within uncertainties. In central collisions the average $\RAA$ of non-strange D mesons is compatible with that of charged particles for $\pt > 8~\gev/c$, while it is larger at lower $\pt$.
The nuclear modification factors for strange and non-strange D mesons are also compared to theoretical models with different implementations of in-medium energy loss.
\end{abstract}

\end{titlepage}
\setcounter{page}{2}

\newpage 

\section{Introduction}
\label{sec:intro}
Ultra-relativistic collisions of heavy nuclei produce a state of strongly-interacting matter characterised by high energy density and temperature.
According to Quantum Chromodynamics (QCD) on the lattice, in these extreme conditions matter undergoes a phase transition to a Quark-Gluon Plasma (QGP) state in which quarks and gluons are deconfined and chiral symmetry is partially restored~\cite{Karsch:2006xs,Borsanyi:2010bp,Borsanyi:2013bia,Bazavov:2011nk}.

Heavy quarks (such as charm and beauty) are predominantly produced  in the early stage of the collision in hard scattering processes between partons of the incoming nuclei. 
Because of their large masses, their production time ($\sim$~0.1 and 0.02~fm/$c$ for charm and beauty, respectively~\cite{Andronic:2015wma}) is shorter than the formation time of the QGP, which is about 0.3-1.5~fm/$c$ at Large Hadron Collider (LHC) energies~\cite{Liu:2012ax}. In contrast, the thermal production and annihilation rates are negligible~\cite{BraunMunzinger:2007tn}. Heavy quarks therefore experience the full evolution of the hot and dense QCD medium. 

During their propagation through the medium, heavy quarks are exposed to interactions with the medium constituents and lose part of their energy via inelastic (gluon radiation)~\cite{Gyulassy:1990ye,Baier:1996sk} or elastic scatterings (collisional processes)~\cite{Thoma:1990fm,Braaten:1991jj,Braaten:1991we}. 
The colour-charge dependence of the strong interaction and parton-mass-dependent effects are predicted 
to influence the amount of energy loss (see~\cite{Andronic:2015wma,Prino:2016cni} for recent reviews).
Low-momentum heavy quarks can participate in the collective expansion of the system as a consequence of multiple interactions with the medium~\cite{Batsouli:2002qf,Greco:2003vf}.
It was also suggested that low-momentum heavy quarks could hadronise
not only via fragmentation in the vacuum, but also via the mechanism of recombination with other quarks in the 
medium~\cite{Greco:2003vf,Andronic:2003zv}. In this scenario, the large abundance of strange quarks in nucleus--nucleus collisions 
with respect to proton--proton collisions is expected to lead to an increased production of $\Ds$ mesons relative to non-strange D mesons~\cite{Kuznetsova:2006hx}. 

The effects of energy loss and the dynamics of heavy-quark hadronisation can be studied using the nuclear modification factor $\Raa$, which compares
 the transverse-momentum ($\pt$) differential production yields in nucleus--nucleus collisions (${\rm d} N_{\rm AA}/{\rm d}\pt$)
with the cross section in proton--proton collisions (${\rm d}\sigma_{\rm pp}/{\rm d}\pt$) scaled by the average nuclear overlap function $\av{T_{\rm AA}}$:
\begin{equation}
\label{eq:Raa}
R_{\rm AA}(\pt)=
{1\over \av{T_{\rm AA}}} \cdot 
{{\rm d} N_{\rm AA}/{\rm d}\pt \over 
{\rm d}\sigma_{\rm pp}/{\rm d}\pt}\,.
\end{equation}

The average nuclear overlap function $\av{T_{\rm AA}}$ is defined as the average number of nucleon-nucleon collisions $\av{N_{\rm coll}}$, which can be estimated via Glauber model calculations~\cite{Glauber:1970jm,Miller:2007ri,Alver:2008aq,Loizides:2014vua}, divided by the inelastic nucleon-nucleon cross section. 

Measurements of prompt D-meson production by the ALICE Collaboration in Pb--Pb collisions at $\sqrtsNN=2.76~\tev$~\cite{ALICE:2012ab,Adam:2015sza,Adam:2015nna,Adam:2015jda} showed a strong suppression of the D-meson yields by a factor of 5--6 for $8 <\pt<12~\gevc$ in the 10$\%$ most central collisions. Recent results from the CMS Collaboration on $\Dzero$ production in the $\pt$ range 2--100 $\gevc$ show a similar suppression for $6 <\pt<10~\gevc$ in the 10$\%$ most central Pb--Pb collisions at $\sqrtsNN=5.02~\tev$, decreasing with increasing $\pt$~\cite{Sirunyan:2017xss}. In contrast, the D-meson nuclear modification factor in p--Pb collisions at $\sqrtsNN=5.02~\tev$, where an extended QGP phase is not expected to be formed, was found to be consistent with unity within uncertainties for $0<\pt <24~\gev/c$~\cite{Adam:2016ich}. These results indicate that the strong suppression is due to substantial final-state interactions of charm quarks with the QGP formed in Pb--Pb collisions. 

In this article, we present the measurement of $\pt$-differential yields and the nuclear modification factor for prompt $\Dzero$, $\Dplus$, $\Dstar$ and $\Ds$ mesons (including their antiparticles), in Pb--Pb collisions at $\sqrtsNN =5.02~\TeV$ collected with the ALICE detector during the LHC Run 2 in 2015. Prompt D mesons are defined as those produced by the hadronisation of charm quarks or from the decay of excited open charm and charmonium states, hence excluding the decays of beauty hadrons. The experimental apparatus is briefly presented in Section~\ref{sec:detector}, together with the data sample used for the analysis. The reconstruction of D-meson hadronic decays and all corrections applied to the raw yields are presented in Section~\ref{sec:analysis}. The procedure used to obtain the proton--proton reference cross section at $\sqrts =5.02~\TeV$ and the estimation of the systematic uncertainties are described in Section~\ref{sec:ppref} and Section~\ref{sec:Syst}, respectively. 
The results for the central (0--10\%), semi-central (30--50\%) and peripheral (60--80\%) collisions are presented in Section~\ref{sec:results}.
A comparison with charged-pion and charged-particle $R_{\rm AA}$ is reported in the same Section, along with detailed comparisons with 
model calculations, including a simultaneous comparison of the $\RAA$ and elliptic flow $v_2$.
Conclusions are drawn in Section~\ref{sec:summary}.

\section{Experimental apparatus and data sample}
\label{sec:detector}

A description of the ALICE experimental apparatus and its performance in pp, p--Pb and Pb--Pb collisions can be found in~\cite{Aamodt:2008zz,Abelev:2014ffa}. The main detectors used in the present analysis are the V0 detector, the Inner Tracking System (ITS)~\cite{Aamodt:2010aa}, the Time Projection Chamber (TPC)~\cite{Alme:2010ke} and the Time Of Flight (TOF) detector~\cite{Akindinov:2013tea}, located inside a large solenoidal magnet providing a uniform magnetic field of 0.5~T parallel to the LHC beam direction ($z$ axis in the ALICE reference system), and the Zero Degree Calorimeters (ZDC)~\cite{PUDDU2007397}, located at $z = \pm 112.5$~m from the nominal interaction point. The analysed sample consists of Pb--Pb collision data recorded with a minimum-bias interaction trigger that required coincident signals in both scintillator arrays of the V0 detector~\cite{Abbas:2013taa}. The V0 detector consists of two scintillator arrays, which cover the full azimuth in the pseudorapidity intervals $-3.7< \eta <-1.7$ and $2.8< \eta <5.1$. Events produced by the interaction of the beams with residual gas in the vacuum pipe were rejected offline using the V0 and the ZDC timing information. Only events with a reconstructed interaction point (primary vertex) within $\pm10$~cm from the centre of the ITS detector along the beam line were used in the analysis. For the data sample considered in this paper, the probability of in-bunch collision pileup (i.e. collisions with two or more simultaneous interactions per bunch crossing) was negligible, while the request of at least a hit in one of the two innermost layers of the ITS rejected tracks produced in out-of-bunch pileup collisions.

Collisions were divided into centrality classes, determined from the sum of the V0 signal amplitudes and defined in terms of percentiles of the hadronic Pb--Pb cross section. In order to relate the centrality classes to the collision geometry, the distribution of the V0 summed amplitudes was fitted with a function based on the Glauber model~\cite{Glauber:1970jm,Miller:2007ri,Alver:2008aq,Loizides:2014vua} combined with a two-component model for particle production~\cite{Abelev:2013qoq}, which decomposes particle production in nucleus-nucleus collisions into the contributions due to soft and hard interactions. The centrality classes used in the present analysis, together with the corresponding average nuclear overlap function $\av{\TAA}$~\cite{ALICE-PUBLIC-2018-011} and the number of events ($N_{\rm events}$) in each class, are summarised in Table~\ref{tab:Nevents}. The corresponding integrated luminosity is about $L_{\rm int}~\approx 13~\mub^{-1}$~\cite{Adam:2015ptt}.

	\begin{table}[!t]
	\centering
	\begin{tabular}{ccc}
	\hline
	Centrality class & $\langle\TAA\rangle$ (mb$^{-1}$)& $N_{\rm events}$\\
	\hline
	\phantom{0}0--10\% & $23.07\pm0.44$ & $10.4 \times 10^6$ \\
	30--50\% & $3.90\pm0.11$ & $20.8 \times 10^6$\\
	60--80\% & $0.417 \pm 0.014$ & $20.8 \times 10^6$\\
	\hline
	\end{tabular}		
	\caption{Average nuclear overlap function and number of events for the three centrality classes used in the analysis.}
	\label{tab:Nevents}
	\end{table}

\section{Data analysis}
\label{sec:analysis}

The D mesons and their charge conjugates were reconstructed in the decay channels $\DtoKpi$ (with branching ratio, BR, of $(3.93 \pm 0.04)\%$), $\DtoKpipi$ (BR of $(9.46 \pm 0.24)\%$), $\DstartoDpi$ (BR of $(67.7\pm 0.5)\%$) and $\Ds\to\phi\pi^+\to {\rm K^+K^-}\pi^+$ (BR of $(2.27 \pm 0.08)\%$)~\cite{Olive:2016xmw}. $\Dzero$, $\Dplus$ and $\Ds$ candidates were defined using pairs and triplets of tracks with proper charge-sign combination having $|\eta| < 0.8$, $\pt> 0.4~\GeV/c$, a minimum number of 70 (out of 159) associated space points in the TPC and at least two hits (out of six) in the ITS, with at least one in the two innermost layers. $\Dstar$ candidates were formed by combining $\Dzero$ candidates with tracks having $|\eta| < 0.8$, $\pt > 0.1~\GeV/c$ and at least three associated hits in the ITS. For $\Ds$ candidate selection, one of the two pairs of opposite-sign tracks was required to have an invariant mass compatible with the $\phi$ mass ($m_{\phi}=1019.461\pm0.019 ~\MeV/c^2$~\cite{Olive:2016xmw}). In particular, the difference between the reconstructed ${\rm K}^+{\rm K}^-$ invariant mass and $\phi$ mass was required to be less than 5--10$~\MeV/c^2$ depending on the $\Ds$ $\pt$ interval. This selection preserves 70--85\% of the $\Ds$ signal. 

The selection of tracks with $|\eta|<0.8$ limits the D-meson acceptance in rapidity, which, depending on $\pt$, varies from $|y|<0.6$ for $\pt=1~\GeV/c$ to $|y|<0.8$ for $\pt > 5~\GeV/c$. 
A $\pt$-dependent fiducial acceptance cut, $|y_{\rm D}| < y_{\rm fid}(\pt)$, was therefore applied to the D-meson rapidity. The value of $y_{\rm fid}(\pt)$ increases from 0.6 to 0.8 in the range $1 < \pt < 5$~GeV/$c$, and the variation can be described according to a second-order polynomial function. For $\pt > 5$~GeV/$c$ one has $y_{\rm fid} = 0.8$.

The selection strategy is similar to the one used in previous analyses~\cite{Abelev:2014ipa, Adam:2015jda} and is mainly based on the separation between primary and secondary vertex, the displacement of the tracks from the primary vertex and the pointing of the reconstructed D-meson momentum to the primary vertex. In comparison to previous analyses, additional selection criteria were exploited. In particular, the normalised difference between the measured and expected transverse-plane impact parameters of each of the decay particles (already introduced in~\cite{Acharya:2017jgo}) and the transverse-plane impact parameter to the primary vertex ($d_0^{xy}$) of the D-meson candidates were used. Besides the rejection of the combinatorial background, a selection based on the latter two variables has the advantage to suppress significantly the fraction of D mesons coming from beauty-hadron decays (feed-down) and hence reduce the associated systematic uncertainty. The cut values on the selection variables were optimised in each centrality class independently, in order to obtain a large statistical significance of the D-meson signals, while keeping the selection efficiency of promptly produced D mesons as large as possible. Further background reduction was obtained by applying particle identification for charged pions and kaons with the TPC and TOF detectors. A $\pm 3\,\sigma$ window around the expected mean values of specific ionisation energy loss ${\rm d}E/{\rm d} x$ in the TPC gas and time-of-flight from the interaction point to the TOF detector was used for the identification, where $\sigma$ is the resolution on these two quantities. In central collisions, a $2\,\sigma$ selection was used for $\Dstar$ and $\Dplus$ (for $\pt < 3~\gev /c$) candidates. For $\Ds$ candidates, tracks without a TOF signal (mostly at low momentum) were identified using only the TPC information and requiring a $2\,\sigma$ compatibility with the expected ${\rm d}E/{\rm d} x$. The stricter PID selection strategy was needed due to the large background of track triplets and, in case of $\Ds$, because of its short lifetime, which limits the effectiveness of the geometrical selections on the displaced decay-vertex topology.

\begin{figure}[!t]
\begin{center}
\includegraphics[width=0.9\textwidth]{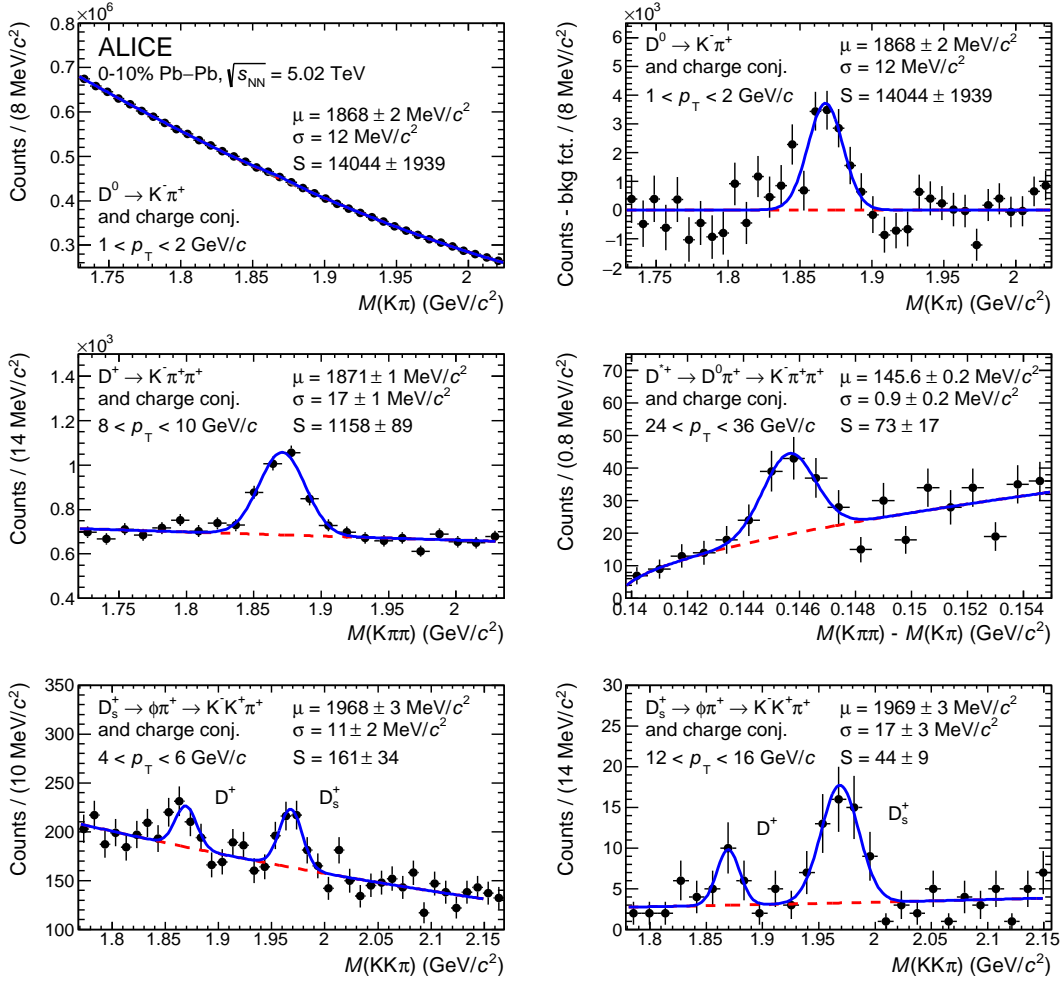}
\caption{Invariant-mass distributions for the four D-meson species in selected $\pt$ intervals for the centrality class 0--10\%. Fitted values for the meson mass $\mu$, width $\sigma$ and raw yield $S$ are also given. 
Top row: $\Dzero$ mesons with $1<\pt<2~\gev/c$, before (left) and after (right) subtraction of the background fit function. For this $\pt$ interval, the width of the Gaussian used to describe the signal is fixed to the value obtained in the simulations.
Middle row: $\Dplus$ mesons with $8<\pt<10~\gev/c$ and $\Dstar$ mesons (difference of $M({\rm K}\pi\pi)$ and $M({\rm K}\pi)$) with $24<\pt<36~\gev/c$.
Bottom row: $\Ds$ mesons with $4<\pt<6~\gev/c$ and $12<\pt<16~\gev/c$; the $\rm \Dplus\to K^+K^-\pi^+$ signal is visible on the left of the $\Ds$ signal. 
}
\label{fig:DInvMass010}
\end{center}
\end{figure}

The $\Dzero$, $\Dplus$ and $\Ds$ raw yields were obtained from binned maximum-likelihood fits to the candidate invariant-mass ($M$) distributions, while for the $\Dstar$ the mass difference $\Delta M = M({\rm K} \pi\pi) -M({\rm K} \pi)$ distributions were used. Examples for these distributions are shown in Fig.~\ref{fig:DInvMass010} for the centrality class 0--10\%. The $\Dzero$, $\Dplus$ and $\Ds$ candidate invariant-mass distributions were fitted with a function composed of a Gaussian term for the signal and an exponential function to describe the background shape, with the exception of the $\Dzero$ $\pt$ intervals 1--2~$\gev/c$ and 2--3~$\gev/c$, where the background was found to be better described by a second-order polynomial function (a fourth-order polynomial was used in 1--2~$\gev/c$ for the 0--10\% centrality class). The $\Delta M$ distribution of $\Dstar$ candidates was fitted with a Gaussian function for the signal and a threshold function multiplied by an exponential for the background ($a\,\sqrt{\Delta M - m_{\pi} } \cdot {\rm e}^{b (\Delta M-m_{\pi})}$, where $m_\pi$ is the pion mass and $a$ and $b$ are free parameters). The contribution of signal candidates that are present in the invariant-mass distribution of the $\Dzero$ meson with the wrong decay-particle mass assignment (reflection), was parametrised by fitting the simulated reflection invariant-mass distributions with a double Gaussian function, and it was included in the total $\Dzero$ fit function. The ratio between the reflected signal and the yields of the $\Dzero$ was taken from simulations (typically 2-5\% of the raw yield, depending on $\pt$)~\cite{Abelev:2014ipa}. The Monte Carlo simulation used for this study is the same one used to determine the reconstruction efficiency, as described in the following dedicated paragraph. In addition, given the critical signal extraction induced by the small signal-to-background ratio of the $\Dzero$ meson in $1<\pt<2~\GeV/c$, the width of the Gaussian used to describe the signal was fixed to the value obtained in the simulations. The Gaussian widths obtained from the simulations were found to be consistent with those extracted from the data in the full $\pt$ range, for all measured centrality classes, with deviations of at most 10--15\%. In the fit to the $\Ds$-candidate invariant-mass distribution, an additional Gaussian was used to describe the $\rm \Dplus\to K^+K^-\pi^+$ signal on the left of the $\Ds$ signal. The statistical significance  $\rm S/\sqrt{S+B}$ of the observed signals, estimated within 3 standard deviations, varies from 5 to 33 depending on the D-meson species, the $\pt$ interval, and the centrality class. 

The D-meson raw yields were corrected in order to obtain the
$\pt$-differential yields of prompt D mesons 

\begin{equation}
  \label{eq:dNdpt}
  \left.\frac{{\rm d} N^{\rm D}}{{\rm d}\pt}\right|_{|y|<0.5}=
  \frac{\left.f_{\rm prompt}(\pt)\cdot \frac{1}{2} N_{\rm raw}^{\rm
        D+\overline{D}}(\pt)\right|_{|y|<y_{\rm fid}(\pt)}}{\Delta\pt \cdot
    \alpha_y(\pt) \cdot ({\rm Acc}\times\epsilon)_{\rm prompt}(\pt)
    \cdot{\rm BR} \cdot N_{\rm events}}\,.
\end{equation}

The raw yields $N_{\rm raw}^{\rm D+\overline{D}}$ were divided by a factor of two to obtain the charge-averaged (particle and antiparticle) yields. To correct for the contribution of feed-down from beauty-hadron decays, the raw yields were multiplied by the fraction of promptly produced D mesons, $f_{\rm prompt}$ (see Eq.~\ref{eq:fcNbMethod}). Furthermore, they were divided by the product of prompt D-meson acceptance and efficiency $({\rm Acc}\times\epsilon)_{\rm prompt}$, by the branching ratio {\rm BR} of the decay channel, by the transverse momentum interval width $\Delta \pt$ and by the number of events $N_{\rm events}$. The $({\rm Acc}\times\epsilon)_{\rm prompt}$ correction includes the tracking efficiency, the acceptance of pions and kaons, and the kinematical and topological selection efficiency of D mesons. The factor $\alpha_y(\pt)=y_{\rm fid}(\pt)/0.5$ normalises the corrected yields measured in $|y|<y_{\rm fid}(\pt)$ to one unit of rapidity $|y|<0.5$, assuming a flat rapidity distribution for D mesons in $|y|<y_{\rm fid}(\pt)$.  This assumption was validated to the 1\% level with simulations for pp collisions~\cite{ALICE:2011aa, Skands:2009zm}
and it is justified also for Pb--Pb collisions. For example, measurements of the prompt and non-prompt J/$\psi$ $\Raa$ in Pb--Pb collisions at $\sqrtsNN=2.76~\TeV$ do not exhibit a significant rapidity dependence~\cite{Khachatryan:2016ypw}.

The correction for acceptance and efficiency $({\rm Acc}\times\epsilon)_{\rm prompt}$ was determined using Monte Carlo simulations with a detailed description of the detector and its response, based on the GEANT3 transport package~\cite{Brun:1994aa}. The underlying Pb--Pb events at $\sqrtsNN = 5.02$~TeV were simulated using the HIJING v1.383 generator~\cite{Wang:1991hta} and D-meson signals were added using the PYTHIA v6.421 generator~\cite{Sjostrand:2006za} with Perugia-2011 tune. Each simulated PYTHIA pp event contained a $\rm c\overline c$ or $\rm b\overline b$ pair, and D mesons were forced to decay into the hadronic channels of interest for the analysis. In the most central event class, the $\pt$ distribution of D mesons in the MC simulation for $\pt>2~\GeV/c$ was weighted in order to match the shape measured in data for \mbox{$\rm D^0$ mesons} in finer $\pt$ intervals with respect to those used in the analysis. In the centrality classes and $\pt$ ranges where an analysis in finer $\pt$ intervals was not possible, the simulated D-meson $\pt$ distribution was weighted to match the shape given by model calculations. In particular, fixed-order plus next-to-leading-log perturbative QCD calculations (FONLL)~\cite{Cacciari:1998it,Cacciari:2001td} multiplied by the $\RAA(\pt)$ of D mesons computed using the BAMPS model (which implements both elastic and radiative processes) for the 30--50\% centrality class~\cite{Uphoff:2011ad,Fochler:2011en,Uphoff:2012gb} were used for the corresponding centrality class. For the $\pt$ intervals 1--2$~\GeV/c$ and 16--50$~\GeV/c$ in the 0--10\% centrality class and for the 60--80\% centrality class, where the $\Raa$ is nearly flat in the measured $\pt$ interval, only the FONLL calculations were used.

\begin{figure}[!t]
\begin{center}
\includegraphics[width=0.34\textwidth]{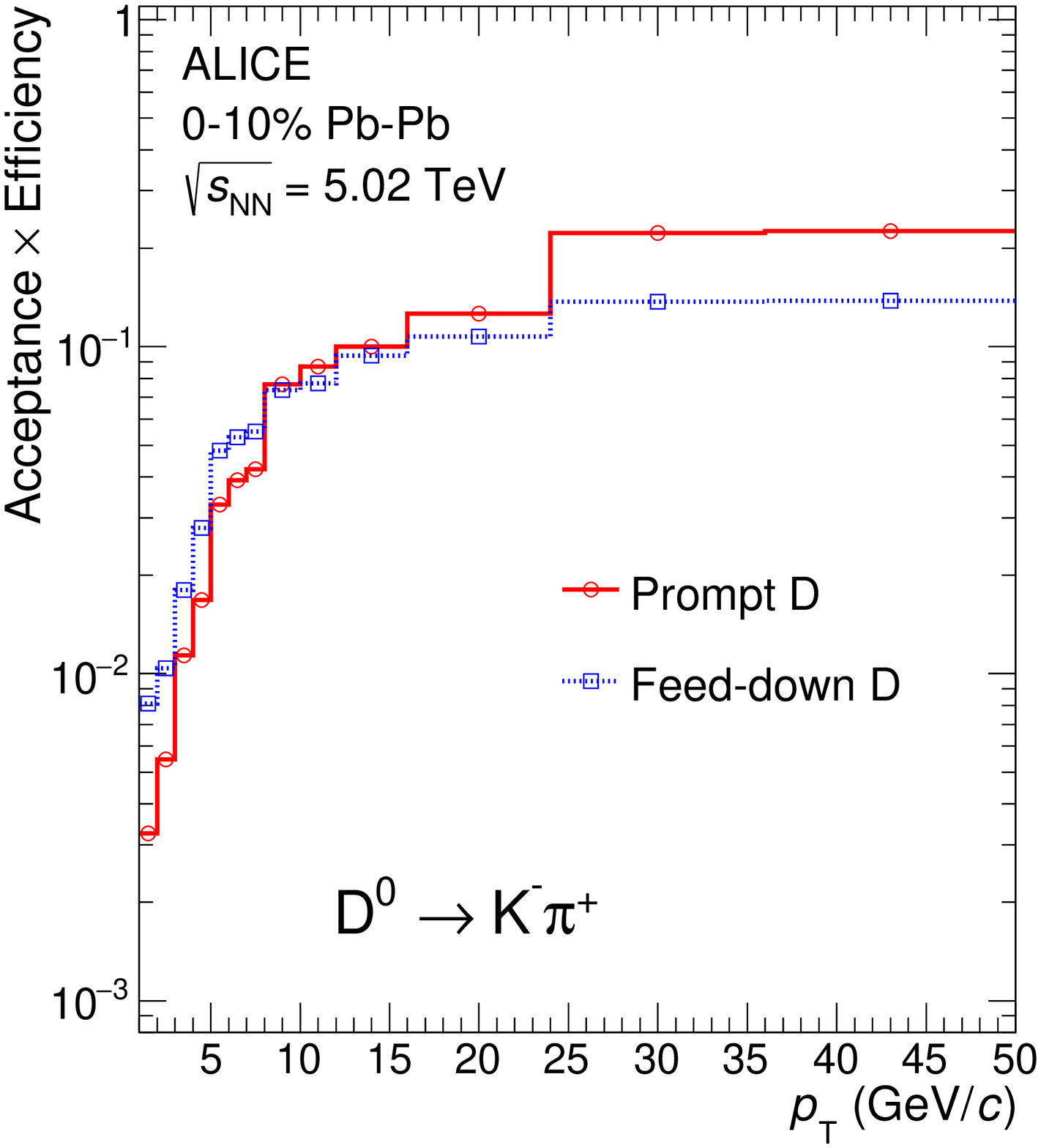}
\includegraphics[width=0.34\textwidth]{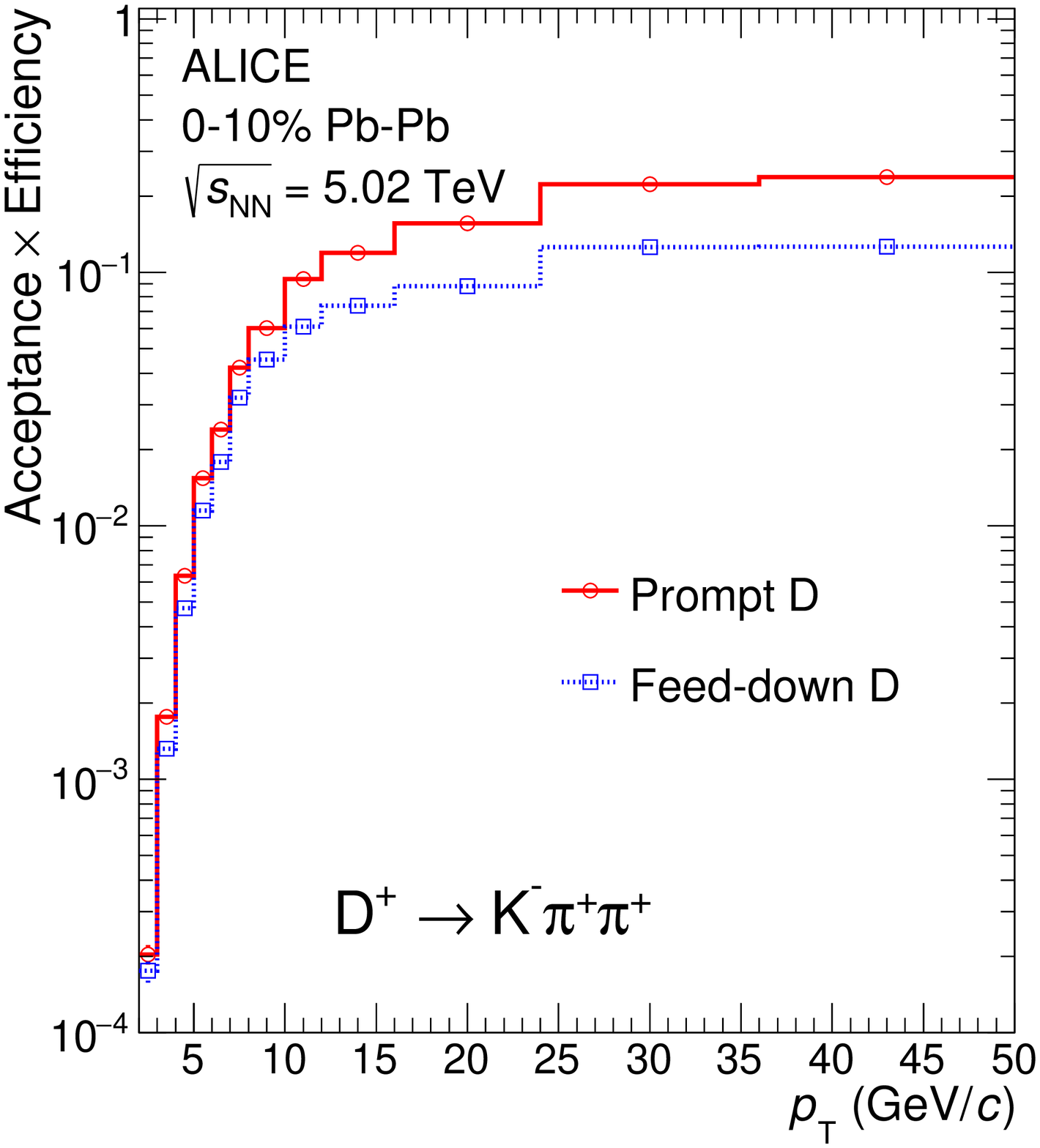}
\includegraphics[width=0.34\textwidth]{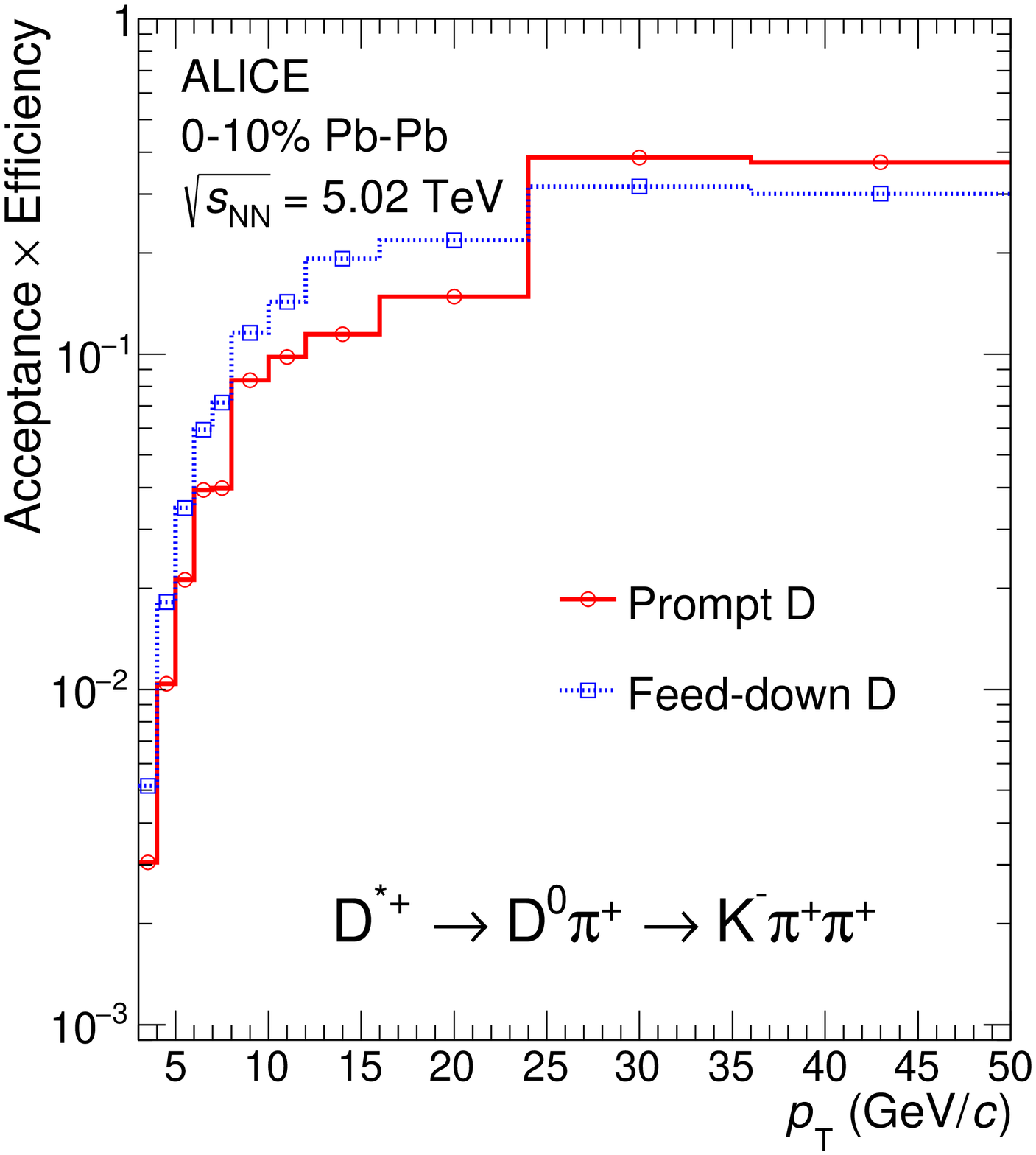}
\includegraphics[width=0.34\textwidth]{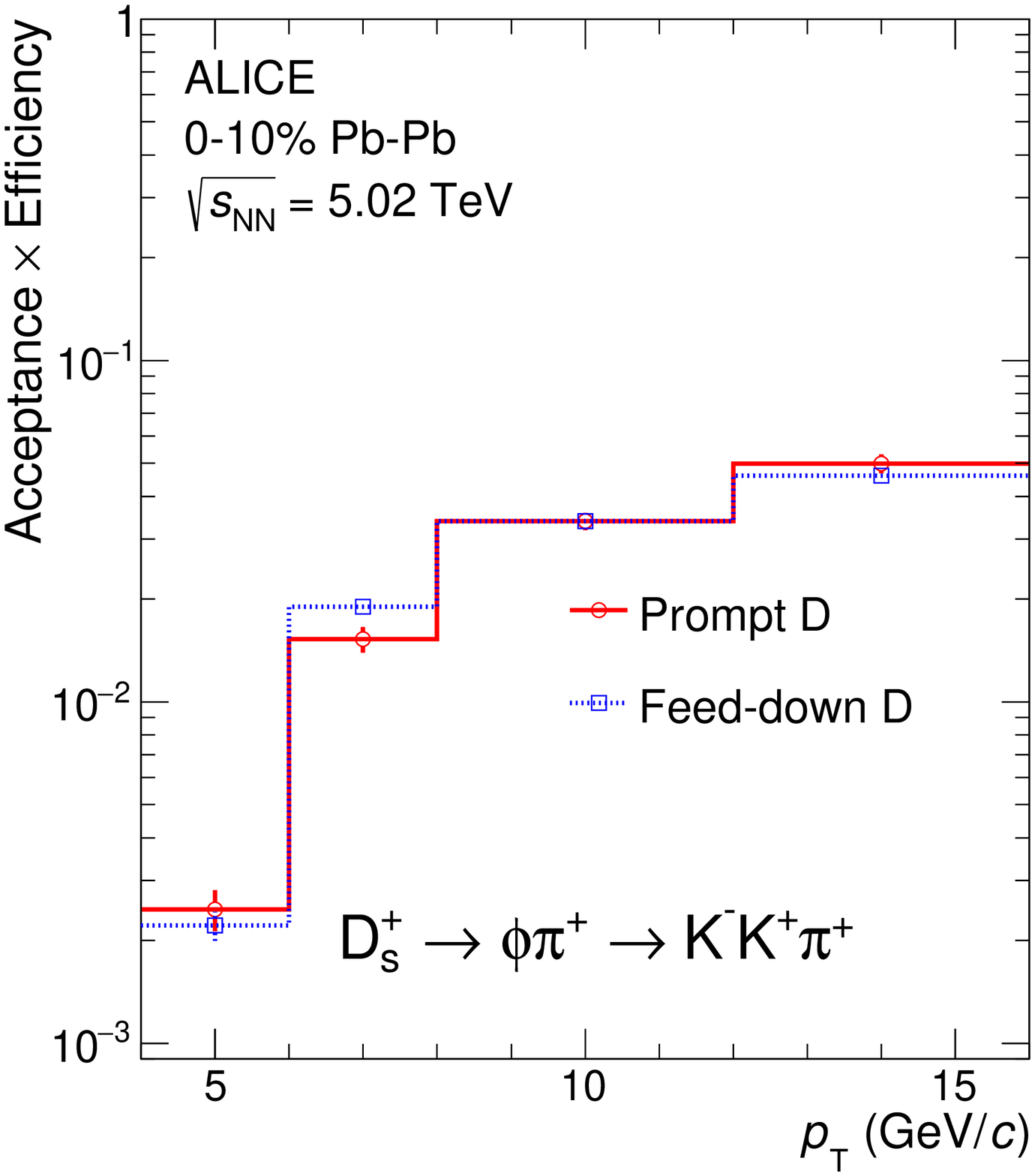}
\caption{Product of acceptance and efficiency as a function of $\pt$ for prompt (red circles) and feed-down (blue squares) D mesons in Pb--Pb collisions for the 0--10\% centrality class obtained from MC simulations. 
} 
\label{fig:Deff}
\end{center}
\end{figure}

Figure~\ref{fig:Deff} shows the acceptance-times-efficiency $({\rm Acc}\times\varepsilon)$ for prompt and feed-down D mesons with rapidity $|y| < y_{\rm fid}(\pt)$ in the centrality class 0--10\%, after the aforementioned $\pt$-distribution weighting procedure. The difference between the $({\rm Acc}\times\varepsilon)$ factor for prompt and feed-down D-mesons arises from the geometrical selections applied, given the different decay topology of D mesons coming from B decays. In particular, the feed-down D mesons are on average more displaced from the primary vertex due to the large B-meson lifetime ($c\tau\approx 500~\mum$~\cite{Olive:2016xmw}) and therefore are more efficiently selected by the majority of the analysis cuts (e.g. for $\Dzero$ and $\Dstar$ in most of the $\pt$ intervals). On the contrary, the selections on the difference between measured and expected decay-track impact parameters and on the D-meson impact parameter reject more feed-down D mesons, thus reducing the feed-down efficiencies as compared to the previous analyses (e.g. for $\Dplus$ and $\Ds$). The $({\rm Acc}\times\varepsilon)$ is higher for more peripheral collisions, by up to a factor larger than two at low $\pt$, since less stringent selections can be applied because of the lower combinatorial background.

The $f_{\rm prompt}$ factor was obtained, following the procedure introduced in~\cite{ALICE:2012ab}, by subtracting the contribution of D mesons from beauty-hadron decays from the measured raw yield in each $\pt$ interval. It was estimated using perturbative QCD calculations, efficiencies from MC simulations, and an hypothesis on the $\RAA$ of feed-down D mesons. The expression for $f_{\rm prompt}$ reads:
\begin{equation}
  \label{eq:fcNbMethod}
\begin{split}
f_{\rm prompt} &= 1-\frac{ N^{\rm D+\overline D\,\textnormal{feed-down}}_{\rm raw}} { N^{\rm D+\overline D}_{\rm raw} }\\
 &= 1 -    \RAA^{\textnormal{feed-down}} \cdot    \langle \TAA \rangle 
	 \cdot \left( \frac{{\rm d} \sigma}{{\rm d}\pt }
         \right)^{\textnormal{FONLL,\,EvtGen}} _{{\textnormal{
               feed-down},\,|y|<0.5}} \cdot
         \frac{\Delta\pt\cdot\alpha_y\cdot({\rm
             Acc}\times\epsilon)_{\textnormal{feed-down}}\cdot {\rm
             BR} \cdot N_{\rm events}  }{ \frac{1}{2} N^{\rm D+\overline D}_{\rm raw}  } \, .
\end{split}
\end{equation}
In this expression, $N^{\rm D+\overline D}_{\rm raw}$ is the measured raw yield and $N^{{\rm D+\overline D}\,\textnormal{feed-down}}_{\rm raw}$ is the estimated raw yield of $\rm D$ mesons from beauty-hadron decays. In detail, the beauty-hadron production cross section in pp collisions at $\sqrt{s}=5.02~\tev$, estimated with FONLL calculations~\cite{Cacciari:2012ny}, was folded with the beauty-hadron${\rightarrow \rm D}+X$ decay kinematics using the EvtGen package~\cite{Lange2001152} and multiplied by $\langle \TAA \rangle$ of the corresponding centrality class, by the $({\rm Acc}\times\varepsilon)$ for feed-down $\rm D$ mesons, and by the other factors introduced in Eq.~(\ref{eq:dNdpt}). In addition, the nuclear modification factor of D mesons from beauty-hadron decays was accounted for. The comparison of the $\RAA$ of prompt D mesons ($\RAA^{\rm prompt}$) at $\sqrtsNN=2.76~\tev$~\cite{Adam:2015nna} with that of $\rm J/\psi$ from B-meson decays~\cite{Khachatryan:2016ypw} at the same energy measured by the CMS Collaboration indicates that prompt charmed hadrons are more suppressed than non-prompt charmed hadrons. 
 The $\Raa$ values differ by a factor of about two in central collisions at a transverse momentum of about $10~\GeV/c$~\cite{Adam:2015nna} and this difference is described by model calculations with parton-mass-dependent energy loss. Therefore, for the centrality classes 0--10\% and 30--50\%, the value $\RAA^{\textnormal{feed-down}}=2\cdot\RAA^{\rm prompt}$ was used to compute the correction for non-strange D mesons with $3<\pt<24~\GeV/c$.
 This hypothesis was varied in the range $1<\RAA^{\textnormal{feed-down}}/\RAA^{\rm prompt}<3$ considering the data uncertainties and model variations to estimate a systematic uncertainty. For $1<\pt<3~\GeV/c$ and $24<\pt<50~\GeV/c$, where model calculations predict a reduced difference between the $\RAA$ values of prompt and non-prompt charm hadrons~\cite{Djordjevic:2015hra,He:2014cla}, the hypothesis $\RAA^{\textnormal{feed-down}}=1.5\cdot\RAA^{\rm prompt}$ was used, with a variation in $1<\RAA^{\textnormal{feed-down}}/\RAA^{\rm prompt}<2$ for the systematic uncertainty. In the case of strange D mesons, effects induced by the in-medium hadronisation and increased abundance of strange quarks could influence the ratio of the $\Raa$ values of prompt and feed-down $\Ds$. Therefore, more conservative central values and variation ranges for the hypothesis were used for $\Ds$ mesons, namely $\RAA^{\textnormal{feed-down}}=\RAA^{\rm prompt}$ and $\frac{1}{3}<\RAA^{\textnormal{feed-down}}/\RAA^{\rm prompt}<3$. For the peripheral class 60--80\%, in which the medium effects are milder, also the difference between charm and beauty mesons is assumed to be reduced: the value $\RAA^{\textnormal{feed-down}}=1.5\cdot\RAA^{\rm prompt}$, varied in the range $1<\RAA^{\textnormal{feed-down}}/\RAA^{\rm prompt}<2$, was used for all D-meson species. The resulting $f_{\rm prompt}$ values, for the central hypotheses on $\RAA^{\textnormal{feed-down}}/\RAA^{\rm prompt}$, range from about 0.80 to 0.95, depending on the D-meson species, centrality class and $\pt$ interval. The systematic uncertainties obtained from the variation of the hypotheses are discussed in Section~\ref{sec:Syst}.

\section{Proton--proton reference for $\RAA$}
\label{sec:ppref}

The $\pt$-differential cross sections of prompt D mesons with $|y|<0.5$ in pp collisions at $\sqrt s=5.02~\tev$,
  used as reference for the nuclear modification factor, were obtained by scaling the measurements at $\sqrt{s}=7~\TeV$~\cite{Acharya:2017jgo} to $\sqrts=5.02~\tev$ with FONLL calculations~\cite{Cacciari:2012ny}. These measurements reach up to $\pt=36~\gev/c$ for $\Dzero$, $24~\gev/c$ for $\Dplus$ and $\Dstar$, and $12~\gev/c$ for $\Ds$ mesons.
The uncertainties on the $\pt$-dependent scaling factor from $\sqrts=7~\tev$ to $\sqrts=5.02~\tev$ were determined by varying
the FONLL parameters (charm-quark mass, factorisation and renormalisation scales) as described in~\cite{Averbeck:2011ga}.
 The uncertainties range from $^{+17}_{-\phantom{1}4}\%$ for $1<\pt<2~\gev/c$ to about $\pm3\%$ for $\pt>10~\gev/c$.

At high D-meson $\pt$ ($36<\pt<50~\gev/c$ for $\Dzero$, $24<\pt<50~\gev/c$ for $\Dplus$ and $\Dstar$, and  $12<\pt<16~\gev/c$ for $\Ds$), the FONLL calculation at $\sqrt s=5.02~\tev$~\cite{Cacciari:2012ny} was used as a reference by scaling the values for each meson species to match the central value of the scaled data at lower $\pt$. This procedure is described in Ref.~\cite{Adam:2015sza}. As an example, the total systematic uncertainties on the pp reference for $\Dzero$ mesons with $36<\pt<50~\gev/c$ is $^{+38}_{-28}\%$.

\section{Systematic uncertainties}
\label{sec:Syst}
Systematic uncertainties on the D-meson yield in Pb--Pb collisions were 
estimated considering the following sources:
(i) extraction of the raw yield from the invariant-mass distributions; 
(ii) track reconstruction efficiency; 
(iii) D-meson selection efficiency; 
(iv) PID efficiency; 
(v) generated D-meson $\pt$ shape in the simulation; 
(vi) subtraction of the feed-down from beauty-hadron decays.
In addition, the uncertainties on the branching ratios~\cite{Olive:2016xmw} were considered. A procedure similar to that described in ~\cite{ALICE:2012ab,Adam:2015sza,Adam:2015nna,Adam:2015jda} and outlined 
in what follows was used to estimate the uncertainties as a function of $\pt$ and centrality. 
The systematic uncertainties on the raw yield extraction were evaluated for
each D-meson species and in each $\pt$ interval by varying 
the lower and upper limits of the fit range, and the background fit function. In addition, the same approach was used with a bin-counting method, in which the signal yield was obtained by integrating the invariant-mass distribution after subtracting the background estimated from a fit to the side-bands. It ranges between 2\% and 15\% depending on the D-meson species and $\pt$ interval. In the case of $\Dzero$, an additional contribution due to signal reflections in the invariant-mass distribution was estimated by varying the ratio of the integral of the reflections over the integral of the signal and the shape of the templates used in the invariant-mass fits. For the $\Dzero$ meson in the interval $1<\pt<2~\GeV/c$, the signal line shape was varied by using Gaussian functions with the widths fixed to $\pm 15\%$ with respect to the value expected from Monte Carlo simulations, based on the deviations between the Gaussian width values observed in data and simulations. For the four D mesons, further checks on the stability of the results were performed by repeating the fits varying the invariant-mass bin width. 

The systematic uncertainty on the track reconstruction efficiency was estimated 
by varying the track-quality selection criteria and by comparing the 
probability to match the TPC tracks to the ITS hits in data and simulation.
The comparison of the matching efficiency in data and simulations was made 
after weighting the relative abundances of primary and secondary particles in 
the simulation to match those observed in data, which were estimated via fits 
to the inclusive track impact parameter distributions.
The estimated uncertainty depends on the D-meson $\pt$ and ranges from 
3\% to 8\% for the two-body decay of $\Dzero$ mesons and from 6\% to 12\% 
for the three-body decays of $\Dplus$, $\Dstar$ and $\Ds$ mesons.

To estimate the uncertainty on the PID selection efficiency, for the three 
non-strange D-meson species the analysis was repeated without PID selection. 
The resulting cross sections were found to be compatible with those obtained 
with the PID selection and therefore no systematic uncertainty was assigned.
For the $\Ds$ meson, the lower signal yield and the larger combinatorial 
background prevented a signal estimation without particle identification. In 
this case, a 3\% uncertainty was estimated by repeating the analysis with a $3\,\sigma$ PID selection, similar 
to that used for non-strange D-mesons for which no systematic effects were observed. This value was also verified by comparing the pion and kaon 
PID selection efficiencies in the data and in the simulation and combining the 
observed differences using the $\Ds$ decay kinematics (for this test, pure pion  
samples were selected using strange hadron decays, while kaon samples in the TPC were obtained
using a tight PID selection in the TOF).

The uncertainty on the D-meson selection efficiency (see Cut efficiency in Table ~\ref{sysunc_yieldtable}) originates from
imperfections in the description of the D-meson kinematic properties
and of the detector resolutions and alignments in the simulation. 
It was estimated by comparing the corrected yields obtained by repeating
the analysis with different sets of selection criteria resulting in a  
significant modification of the efficiencies, raw yield and background values. 
The assigned uncertainty for non-strange D mesons is 5\% in most of the $\pt$ 
intervals and it increases to 10--15\% in the lowest $\pt$ intervals, where the efficiencies are 
low and vary steeply with $\pt$, because of the tighter selections. 
A larger uncertainty of 10\% in all $\pt$ intervals was
estimated for $\Ds$ mesons, for which more stringent selection criteria
were utilized in the analysis as compared to non-strange D mesons.

The systematic effect on the efficiency due to a possible difference between 
the real and simulated D-meson transverse momentum distributions
was estimated by using alternative D-meson $\pt$ distributions.
In particular, the $\pt$ distributions from FONLL calculations with
and without hot-medium effects parametrised based on the $\RAA$ in central collisions from the 
BAMPS~\cite{Uphoff:2014hza}, LBT~\cite{Cao:2017hhk} and TAMU~\cite{He:2014cla} models were used in this study.
The uncertainty, which also includes the effect of 
the $\pt$ dependence of the nuclear modification factor, was estimated to be, for non-strange D mesons in central collisions,
about 10\% in the lowest $\pt$ intervals and decreasing to zero for $\pt>5~\gev/c$. For $\Ds$ mesons 
the uncertainty was estimated as 7\% in 4--6~$\gev/c$, 2\% in 6--8~$\gev/c$ and 1\% at higher $\pt$.

The systematic uncertainty on the subtraction of feed-down from beauty-hadron decays 
(i.e.\,the calculation of the $f_{\rm prompt}$ fraction) was estimated by varying i)
the $\pt$-differential feed-down D-meson cross section from the 
FONLL calculation within the theoretical uncertainties, ii)  the ratio of the feed-down and prompt D-meson $\Raa$ in the 
ranges described at the end of Section~\ref{sec:analysis}. The resulting uncertainty ranges between 2\% and 15\%, depending on D-meson species, centrality classes and $\pt$ intervals.

The systematic uncertainties on the $\pt$-differential spectra and $\RAA$ in the two extreme centrality classes are listed for all D-meson 
species in Table~\ref{sysunc_yieldtable} for the lowest $\pt$ interval accessible as well as for the intermediate 
range $7<\pt<8~\gev/c$ ($6<\pt<8~\gev/c$ for the $\Ds$ meson).

\begin{table}[!t]
\centering
\renewcommand{\arraystretch}{1.3}
\scalebox{0.97}{ 
\begin{tabular}{|l|l|cc|cc|cc|cc|}
\hline
\multicolumn{2}{|c|}{Particle}              & \multicolumn{2}{c|}{$\Dzero$}                                             & \multicolumn{2}{c|}{$\Dplus$}            & \multicolumn{2}{c|}{$\Dstar$}                                             & \multicolumn{2}{c|}{$\Ds$}                                             \\ \hline
\multicolumn{10}{|c|}{0--10\% centrality class}                                                                                                                                                                                                                                                      \\ \hline
\multicolumn{2}{|c|}{$\pt$ interval ($\gev/c$)}           & 1--2                  & 7--8                                  & 2--3                  & 7--8  & 3--4                  & 7--8                               & 4--6                  & 6--8                                  \\ \hline
\multicolumn{2}{|l|}{\multirow{2}{*}{Syst. on $\d N/\d \pt$ in Pb--Pb}}      & \multirow{2}{*}{$^{+21}_{-22}$\% }                   &\multirow{2}{*}{ $^{+16}_{-17}$\%}                                         &\multirow{2}{*}{ $22$\% }                   &\multirow{2}{*}{ $^{+16}_{-17}$\%}                        & \multirow{2}{*}{$21$\%}                    &\multirow{2}{*}{ $20$\%}                                       & \multirow{2}{*}{$^{+23}_{-25}$\%  }                  & \multirow{2}{*}{$^{+19}_{-23}$\%   }                                  \\
\multicolumn{2}{|l|}{} & & & & & & & &    \\
\multicolumn{2}{|l|}{{ }{ }{ }{ }Yield extraction}      &15\%                    & 5\%                                     & 12\%                   & 7\%       & 11\%                    & 7\%                                        & 6\%                  & 6\%                                       \\
\multicolumn{2}{|l|}{{ }{ }{ }{ }Tracking efficiency}   & 6\%                    & 7\%                                      & 8.5\%                    & 11\%       & 10\%                    & 10\%                                  & 11\%                    & 12\%                                        \\
\multicolumn{2}{|l|}{{ }{ }{ }{ }PID efficiency}    & 0                   & 0                                        & 0                    & 0  & 0                     & 0                                     & 3\%                    & 3\%                                        \\
\multicolumn{2}{|l|}{{ }{ }{ }{ }Cut efficiency}        & 10\%                    & 6\%                                   & 12\%                    & 8\%       & 13\%                   & 10\%                                      & 13\%                     & 10\%                                        \\
\multicolumn{2}{|l|}{{ }{ }{ }{ }MC $\pt$ shape}        & 8\%                   & 0                                         & 10\%                   & 0       & 4\%                    & 0                                       & 7\%                    & 2\%                                         \\
\multicolumn{2}{|l|}{{ }{ }{ }{ }Branching ratio} & 1.0\%                    & 1.0\%                                       & 2.5\%                    & 2.5\%        & 1.3\%                    & 1.3\%                                       & 3.5\%                   & 3.5\%                      \\
\multicolumn{2}{|l|}{\multirow{2}{*}{{ }{ }{ }{ }Feed-down subtraction}}& \multirow{2}{*}{$^{+6.8}_{-7.3}$\%}                   & \multirow{2}{*}{$^{+12.4}_{-12.8}$\%}                      & \multirow{2}{*}{$^{+2.7}_{-3.0}$\%}                   & \multirow{2}{*}{$^{+6.0}_{-6.3}$\% }      &\multirow{2}{*}{ $^{+6.1}_{-6.5}$\%}                   & \multirow{2}{*}{$^{+11.5}_{-11.8}$\% }                   & \multirow{2}{*}{$^{+4.0}_{-9.5}$\% }                   & \multirow{2}{*}{$^{+6.7}_{-14.7}$\%}                                       \\ 
\multicolumn{2}{|l|}{} & & & & & & & & \\
\cline{3-10}
\multicolumn{2}{|l|}{{ }{ }{ }{ }Centrality limit}      & \multicolumn{8}{c|}{$<$0.1\%} \\ \hline
\multicolumn{2}{|c|}{Syst. on $\d N/\d \pt$ in pp and} 
        &\multirow{2}{*}{$^{+8.8}_{-9.0}$\%}                   
	&\multirow{2}{*}{$^{+8.4}_{-9.4}$\%}   
                 
	&\multirow{2}{*}{$13$\%}                   
	&\multirow{2}{*}{$^{+8.8}_{-9.1}$\%}   

	&\multirow{2}{*}{$8.3$\%}                   
	&\multirow{2}{*}{$^{+8.1}_{-8.4}$\%}   
                  
	&\multirow{2}{*}{$^{+13}_{-14}$\%}       
	&\multirow{2}{*}{$^{+13}_{-14}$\%}                     \\ 
\multicolumn{2}{|l|}{$\sqrt s$-scaling of the pp ref.} & & & & & & & & \\
\hline
\multicolumn{2}{|c|}{\multirow{2}{*}{Syst. on $\RAA$}} & \multirow{2}{*}{$^{+22}_{-27}$\% }                  & \multirow{2}{*}{$^{+17}_{-16}$\%}                       &\multirow{2}{*}{ $^{+26}_{-27}$\%}                   & \multirow{2}{*}{$19$\%}       & \multirow{2}{*}{$23$\% }                   & \multirow{2}{*}{$21$\% }                    & \multirow{2}{*}{$^{+27}_{-28}$\% }                   & \multirow{2}{*}{$^{+23}_{-26}$\% }                                                     \\
\multicolumn{2}{|l|}{} & & & & & & & & \\ \hline
\multicolumn{10}{|c|}{60--80\% centrality class}                                                                                                                                                                                                                                                      \\ \hline
\multicolumn{2}{|c|}{$\pt$ interval ($\gev/c$)}           & 1--2                  & 7--8                                   & 2--3                  & 7--8  & 1--2                  & 7--8                                  & 2--4                  & 6--8                                   \\ \hline
\multicolumn{2}{|l|}{\multirow{2}{*}{Syst. on $\d N/\d \pt$ in Pb--Pb}}        & \multirow{2}{*}{$22$\%}                    &\multirow{2}{*}{ $^{+12}_{-13}$\%}                                        & \multirow{2}{*}{$12$\% }                   & \multirow{2}{*}{$13$\% }                      & \multirow{2}{*}{$23$\% }                   & \multirow{2}{*}{$14$\% }                                       & \multirow{2}{*}{$23$\% }                   &\multirow{2}{*}{ $20$\%  }                              \\
\multicolumn{2}{|l|}{} & & & & & & & & \\
\multicolumn{2}{|l|}{{ }{ }{ }{ }Yield extraction}      & 10\%                    & 4.5\%                                  & 4\%                    & 3\%      & 13\%                    & 2\%                                  & 10\%                    & 6\%                                       \\
\multicolumn{2}{|l|}{{ }{ }{ }{ }Tracking efficiency}   & 6\%                    & 7\%                                   & 8.5\%                  & 11\%      & 9\%                    & 9\%                                       & 8.5\%                  & 12\%                                         \\
\multicolumn{2}{|l|}{{ }{ }{ }{ }PID efficiency}    & 0                  & 0                                   & 0                   & 0     & 0                    & 0                                       & 3\%                   & 3\%                                      \\
\multicolumn{2}{|l|}{{ }{ }{ }{ }Cut efficiency}        & 10\%                    & 5\%                                     & 6\%                    & 5\%        & 15\%                    & 8\%                                     & 14\%                    & 12\%                                      \\
\multicolumn{2}{|l|}{{ }{ }{ }{ }MC $\pt$ shape}        & 12\%                    & 0                                       & 4\%                    & 0        & 5\%                    & 0                                       & 6\%                    & 2\%                                     \\
\multicolumn{2}{|l|}{{ }{ }{ }{ }Branching ratio} & 1.0\%                    & 1.0\%                                        & 2.5\%                    & 2.5\%        & 1.3\%                    & 1.3\%                                         & 3.5\%                   & 3.5\%                      \\
\multicolumn{2}{|l|}{\multirow{2}{*}{{ }{ }{ }{ }Feed-down subtraction}} & \multirow{2}{*}{$^{+9.0}_{-9.7}$\%}                   & \multirow{2}{*}{$^{+6.1}_{-7.2}$\%}                       & \multirow{2}{*}{$^{+3.0}_{-3.3}$\%}                   & \multirow{2}{*}{$^{+3.8}_{-4.4}$\%}        & \multirow{2}{*}{$^{+4.4}_{-4.8}$\% }                  &\multirow{2}{*}{ $^{+6.4}_{-7.4}$\% }                  & \multirow{2}{*}{$^{+7.2}_{-7.9}$\%}                    & \multirow{2}{*}{$^{+9.2}_{-10.6}$\%}                                      \\ 
\multicolumn{2}{|l|}{} & & & & & & & & \\
	\cline{3-10}
\multicolumn{2}{|l|}{{ }{ }{ }{ }Centrality limit}      & \multicolumn{8}{c|}{3.0\%}    \\ \hline
\multicolumn{2}{|c|}{Syst. on $\d N/\d \pt$ in pp and} 
	& \multirow{2}{*}{$^{+8.8}_{-9.0}$\%}                   
	& \multirow{2}{*}{$^{+8.4}_{-9.4}$\%}   
                 
	& \multirow{2}{*}{$13$\%}                   
	& \multirow{2}{*}{$^{+8.8}_{-9.1}$\%}   
 
	& \multirow{2}{*}{$12$\%}                   
	& \multirow{2}{*}{$^{+8.1}_{-8.4}$\%}   
           
	& \multirow{2}{*}{$13$\%}      
	& \multirow{2}{*}{$^{+13}_{-14}$\%}                     
                                   \\ 
\multicolumn{2}{|l|}{$\sqrt s$-scaling of the pp ref.} & & & & & & & & \\
\hline
\multicolumn{2}{|c|}{\multirow{2}{*}{Syst. on $\RAA$}} & \multirow{2}{*}{$^{+23}_{-28}$\%  }                 & \multirow{2}{*}{$14$\%}                   &\multirow{2}{*}{ $^{+18}_{-20}$\% }                  & \multirow{2}{*}{$16$\%}     &\multirow{2}{*}{ $^{+26}_{-31}$\%}                   &\multirow{2}{*}{ $16$\% }                     & \multirow{2}{*}{$26$\% }                   & \multirow{2}{*}{$23$\% }                                                        \\                                                 
\multicolumn{2}{|l|}{} & & & & & & & & \\\hline
\end{tabular}
}
\caption{Relative systematic uncertainties on the $\d N/\d \pt$ in Pb--Pb collisions, on the extrapolated $\d N/\d \pt$ in pp collisions and on the $\RAA$ of $\Dzero$, $\Dstar$, $\Dplus$, and $\Ds$ in two centrality classes considered in the analysis for the lowest accessible $\pt$ intervals and for the intermediate range $7<\pt<8~\gev/c$ ($6<\pt<8~\gev/c$ for the $\Ds$ meson).}
\label{sysunc_yieldtable}
\end{table}

The systematic uncertainties on the $\Raa$ measurement include those 
on the D-meson corrected yields described above, those on the
proton--proton reference cross section, and the uncertainties on the
average nuclear overlap function. 

The systematic uncertainty on the pp reference used for the calculation of $\RAA$ 
has two contributions. The first one is the systematic uncertainty on the measured 
$\pt$-differential D-meson cross section at $\sqrt s=7$~TeV.
The second contribution is the scaling to 
$\sqrt s = 5.02$~TeV, which has been discussed in Section~\ref{sec:ppref}.

In the calculation of the nuclear modification factor, the systematic 
uncertainty on the feed-down subtraction deriving from
the variation of the parameters of the FONLL calculation was considered to be
correlated in the $\PbPb$ and pp measurements, while all the other sources of systematic uncertainties were treated as uncorrelated.

The uncertainties on the $\RAA$ normalisation are the quadratic sum of 
(i) the pp normalisation uncertainty  (3.5\%), 
(ii) the uncertainty on $\langle T_{\rm AA} \rangle$, which ranges from 1.9\% to 3.4\% depending on the centrality, and
(iii) the variation of raw yield ($<0.1\%$, $2\%$ and $3\%$ for the 0--10\%, 30--50\% 
and 60--80\% centrality classes, respectively) obtained when the centrality intervals are varied to account for the uncertainty on 
the fraction of the hadronic cross section used in the 
Glauber fit to determine the centrality~\cite{Adam:2015sza}, and the branching ratio uncertainty cancels out in the 
ratio.

\section{Results}
\label{sec:results}

\begin{figure}[!t]
 \begin{center}
\subfigure[]{
\includegraphics[angle=0, width=7.5cm]{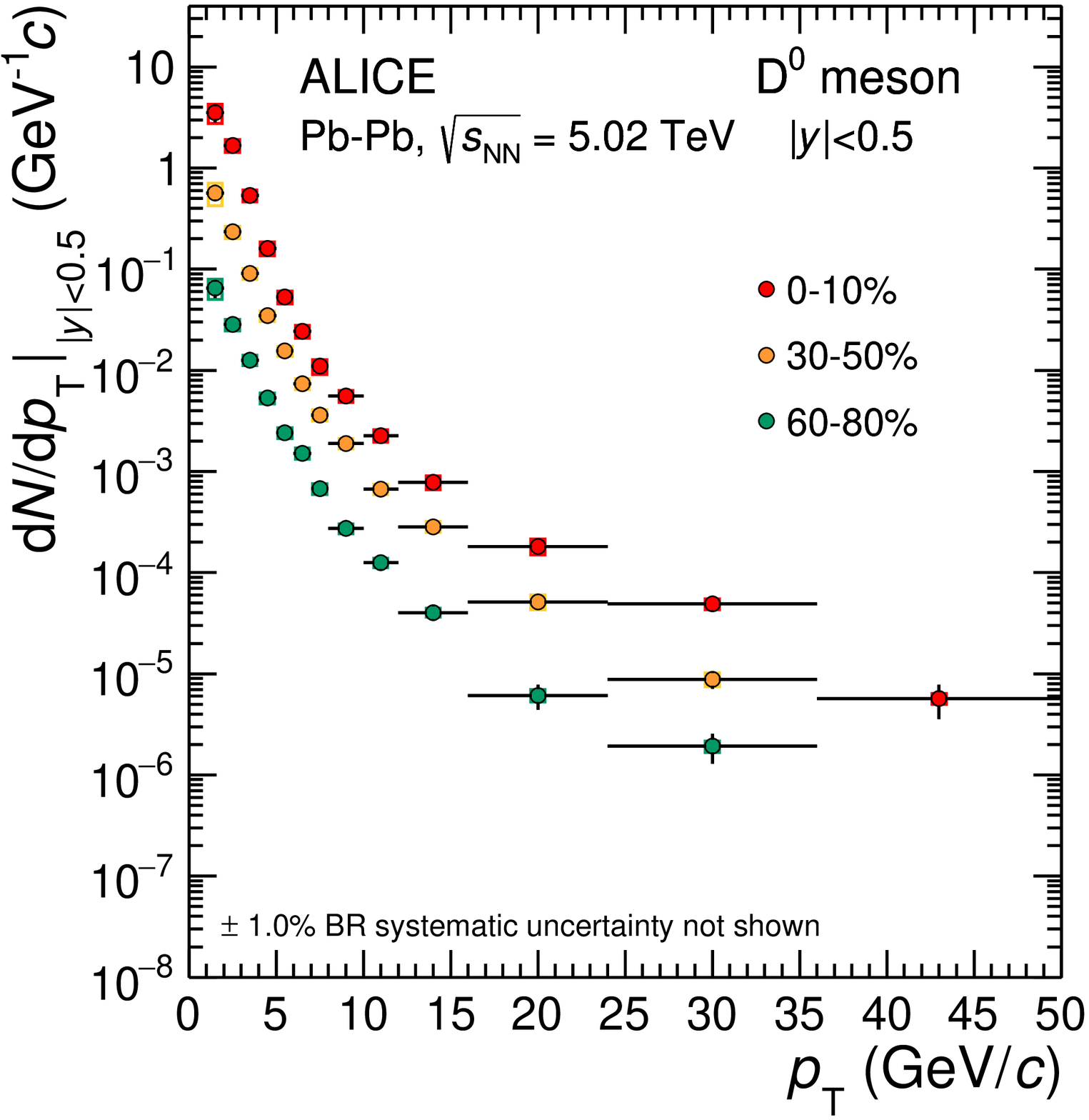}
\label{D0yield}
}
\subfigure[]{
\includegraphics[angle=0, width=7.5cm]{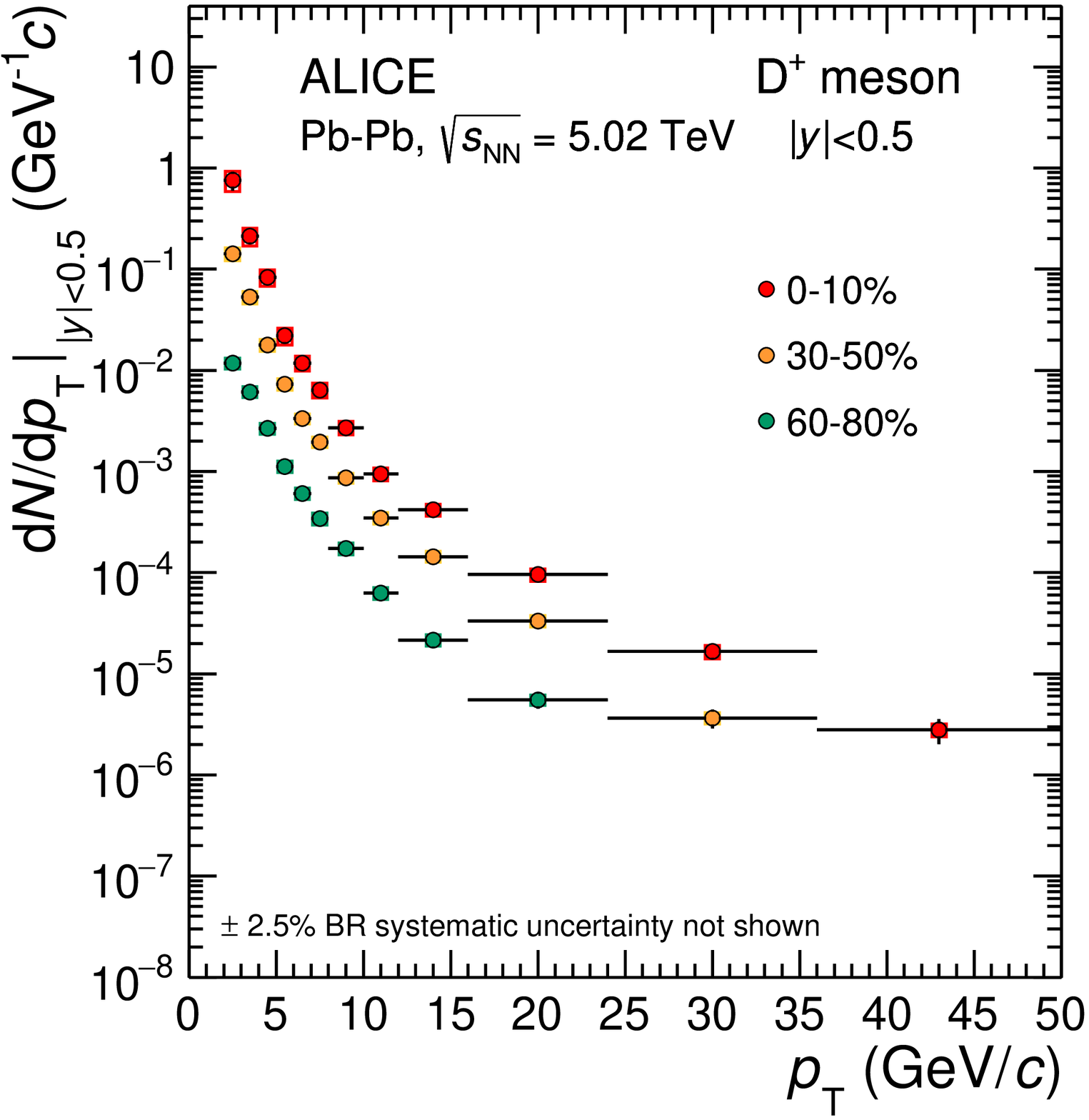}
\label{Dplusyield}
} 
\subfigure[]{
\includegraphics[angle=0, width=7.5cm]{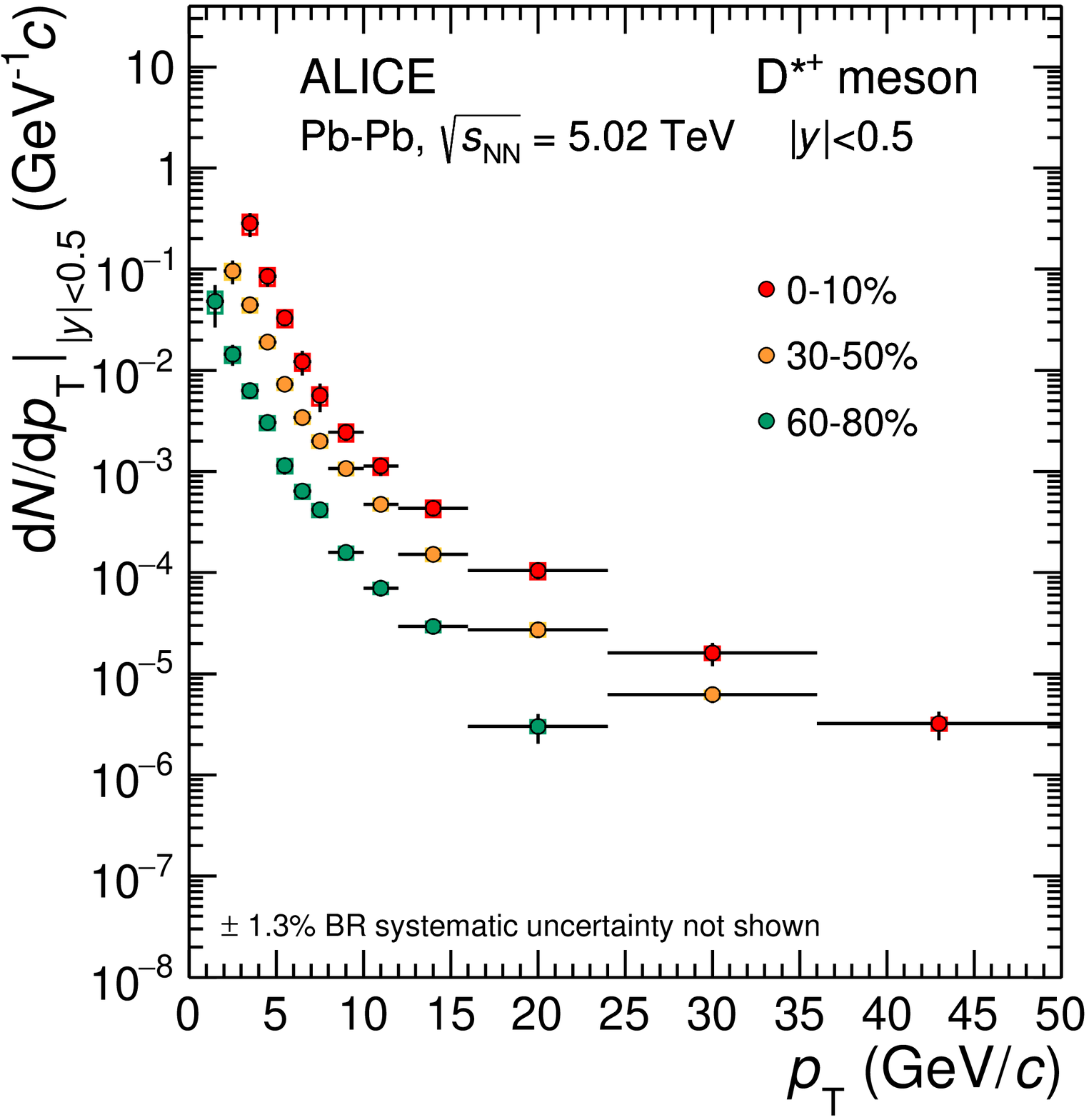}
\label{Dstaryield}
} 
\subfigure[]{
\includegraphics[angle=0, width=7.5cm]{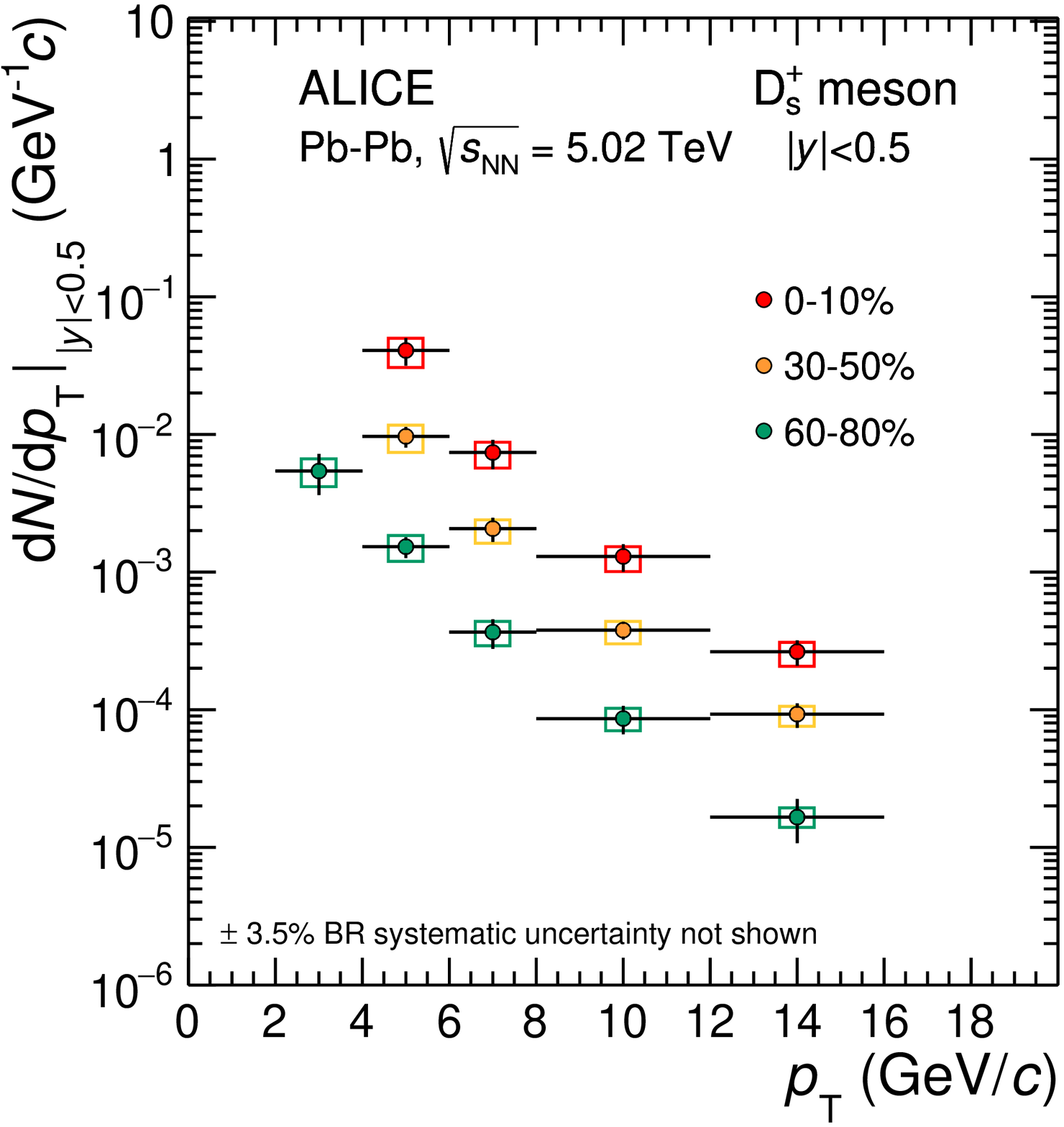}
 \label{Dsyield}
}
 \end{center}
 \caption{Transverse momentum distributions $\d N/\d\pt$ of 
prompt $\Dzero$ (a), $\Dplus$ (b), $\Dstar$ (c) and $\Ds$ (d)  mesons in the 0--10\%, 30--50\% and 60--80\% 
centrality classes in $\PbPb$ collisions 
at $\sqrtsNN=5.02~\tev$. 
Statistical uncertainties (bars) and systematic uncertainties (boxes) are shown. 
The uncertainty on the branching ratios is quoted separately.
Horizontal bars represent bin widths, symbols are placed at the centre of 
the bin. 
}
 \label{fig:DmesCorrYields010} 
\end{figure}

The transverse-momentum distributions $\d N/\d \pt$ of prompt $\Dzero$,
$\Dplus$, $\Dstar$ and $\Ds$ mesons are shown in Fig.~\ref{fig:DmesCorrYields010}
for the 0--10\%, 30--50\% and 60--80\% centrality classes. 
The vertical bars represent the statistical uncertainties and the empty boxes
the systematic uncertainties. 
The uncertainty on the branching ratios is quoted separately.

\begin{figure}[!t]
 \begin{center}
\includegraphics[angle=0, width=0.49\textwidth]{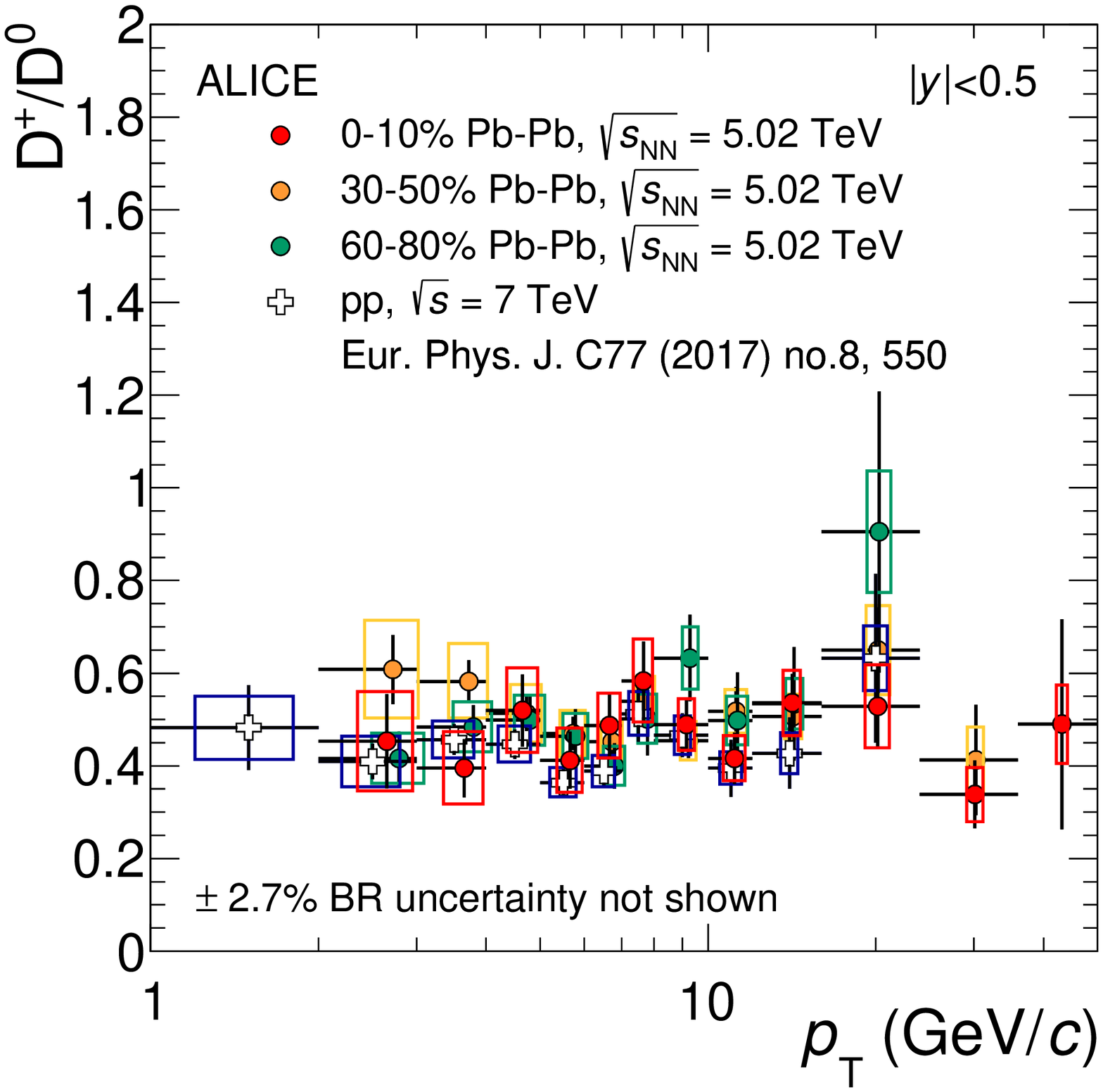}
\includegraphics[angle=0, width=0.49\textwidth]{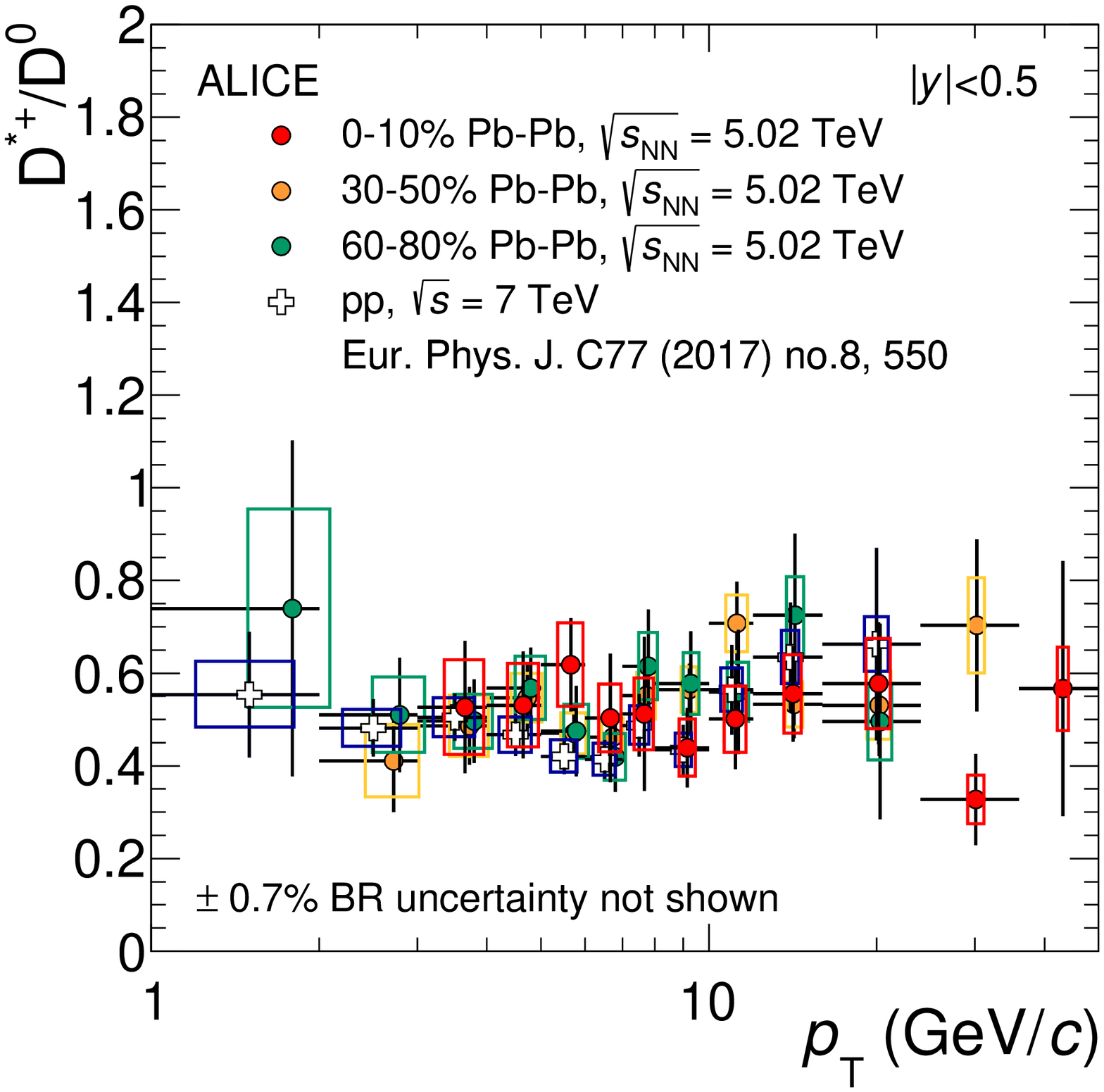}
\includegraphics[angle=0, width=0.49\textwidth]{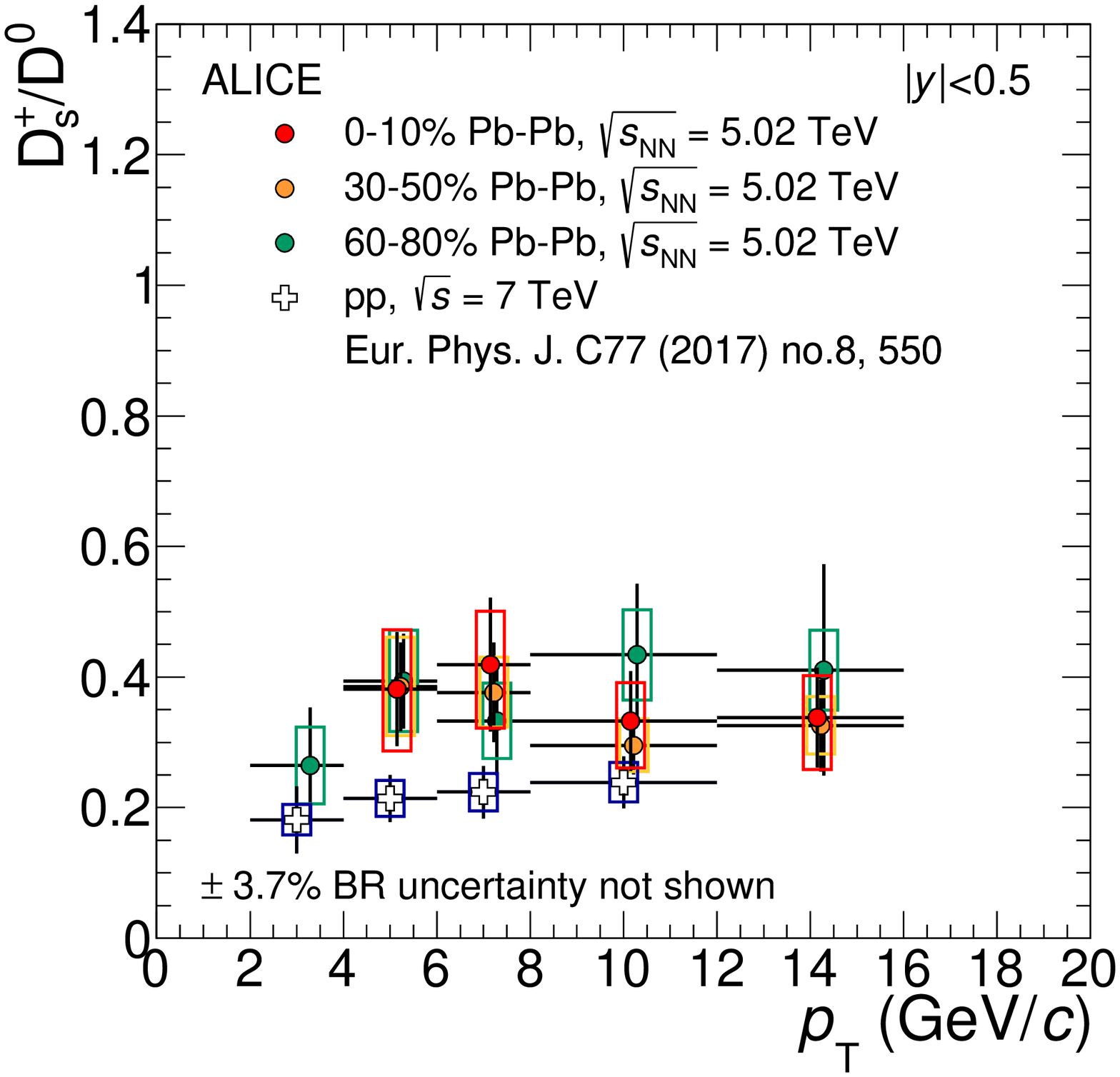}
\includegraphics[angle=0, width=0.49\textwidth]{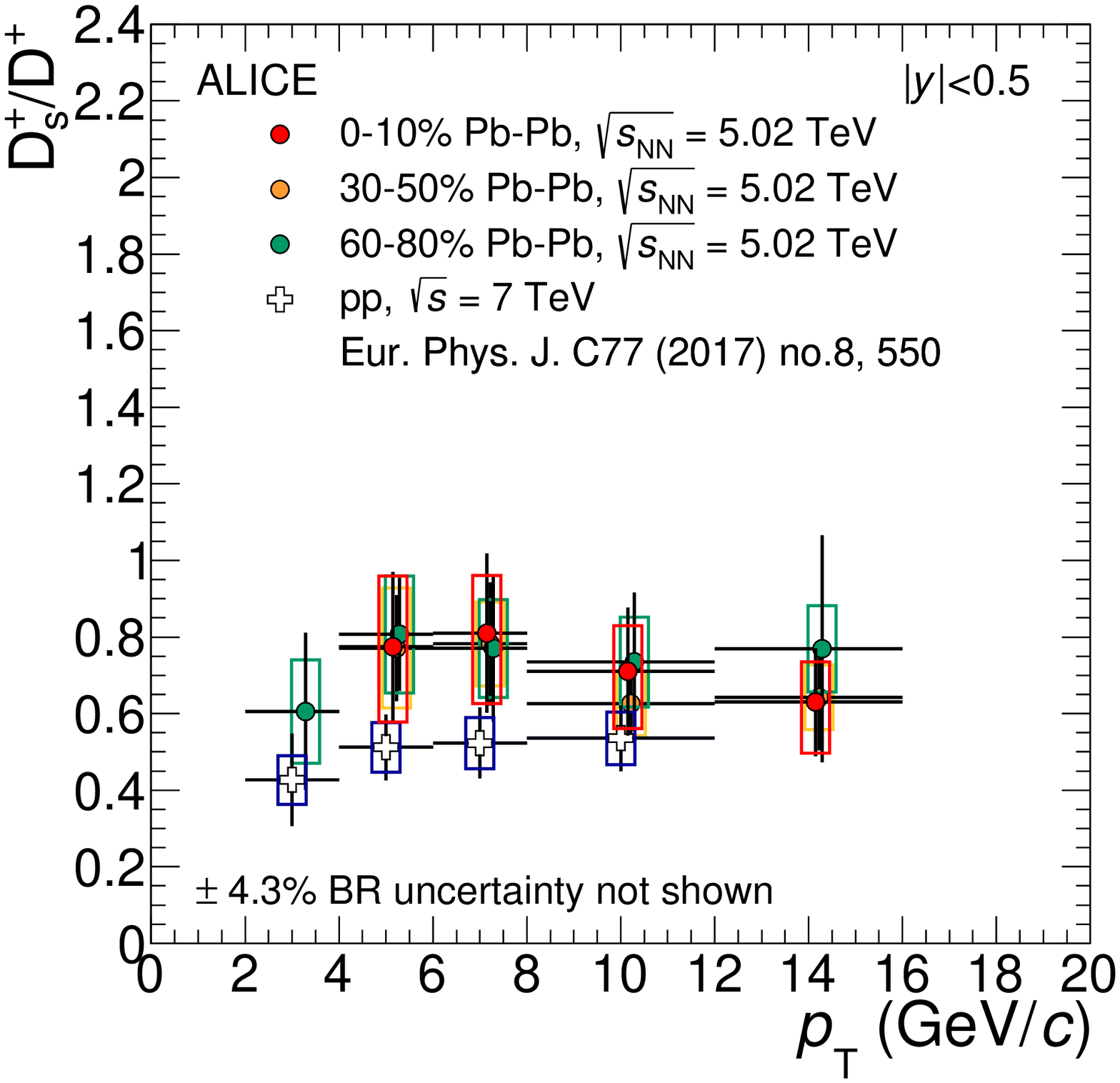}
 \end{center}
 \caption{Ratio of prompt D-meson yields as a function of $\pt$.
Statistical (bars) and systematic (boxes) uncertainties are shown.}
 \label{DmesRatio} 
\end{figure}

Figure~\ref{DmesRatio} shows the $\pt$-dependent ratios of meson yields,
$\Dplus$/$\Dzero$, $\Dstar$/$\Dzero$, $\Ds/\Dzero$ and $\Ds/\Dplus$, 
compared to the values measured in pp collisions at $\sqrt s=7~\tev$~\cite{Acharya:2017jgo}. 
The systematic uncertainties were propagated to the ratios,
considering the contribution from the tracking efficiency as a fully correlated uncertainty among the four D-meson species. The beauty-hadron feed-down subtraction was considered 
as fully correlated among the three non-strange D-meson species, while uncorrelated between $\Ds$ and non-strange D mesons. 
The $\Dplus$/$\Dzero$ and $\Dstar$/$\Dzero$ ratios are compatible in Pb--Pb and pp collisions, 
indicating no significant modification of their relative abundances as a function of $\pt$ and in centrality classes. 
The $\Ds/\Dzero$ and $\Ds/\Dplus$ ratios are measured at $\sqrtsNN=5.02$~TeV with a precision better by a factor about two with respect to 2.76~TeV~\cite{Adam:2015sza}.
The values of these ratios are larger in Pb--Pb than in pp collisions, in all three centrality classes, however
the measurements in the two systems are compatible within about one standard deviation of the combined uncertainties.

\begin{figure}[!h]
 \begin{center}
\includegraphics[angle=0, width=0.43\textwidth]{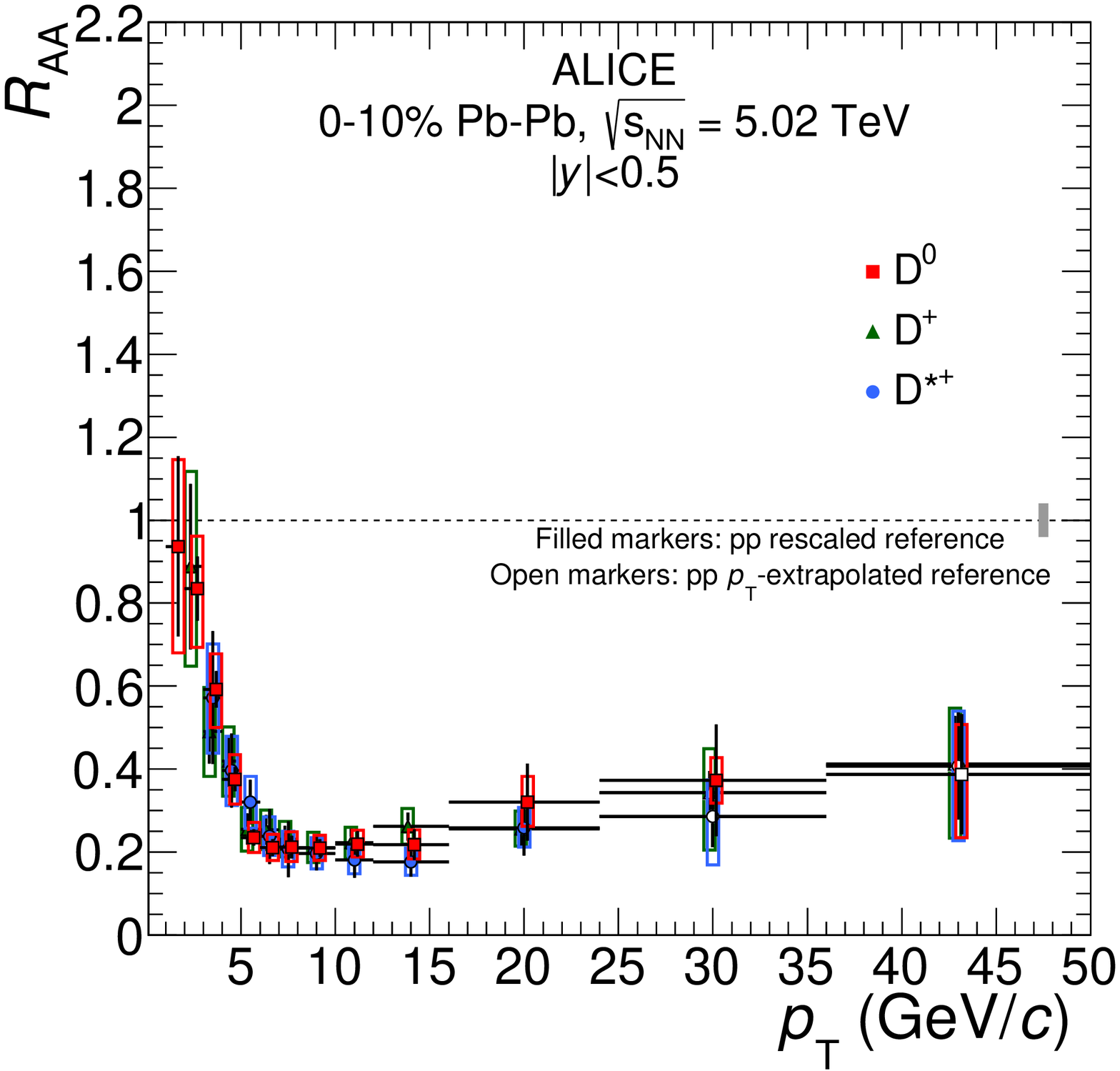}
\includegraphics[angle=0, width=0.43\textwidth]{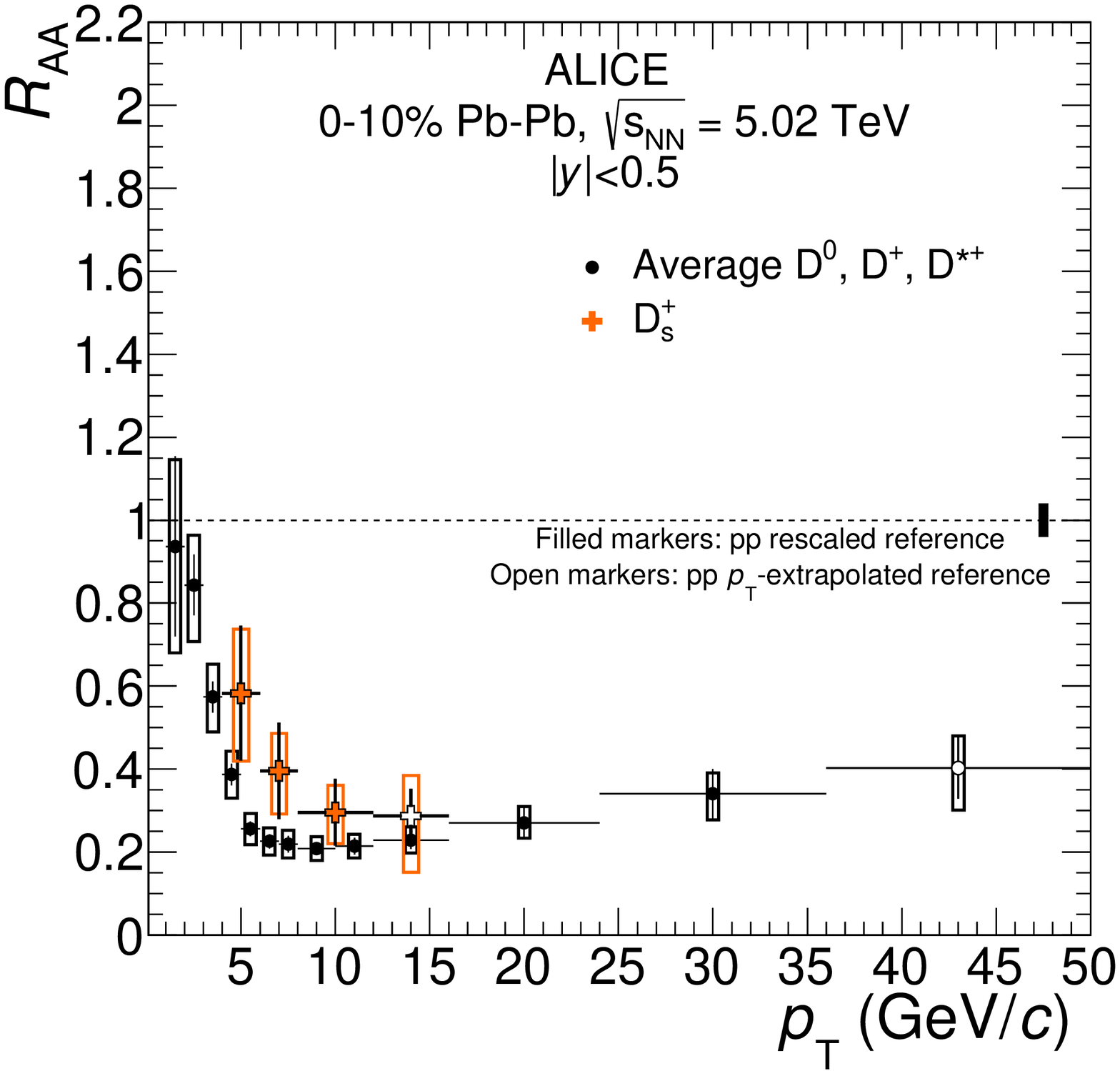}
\includegraphics[angle=0, width=0.43\textwidth]{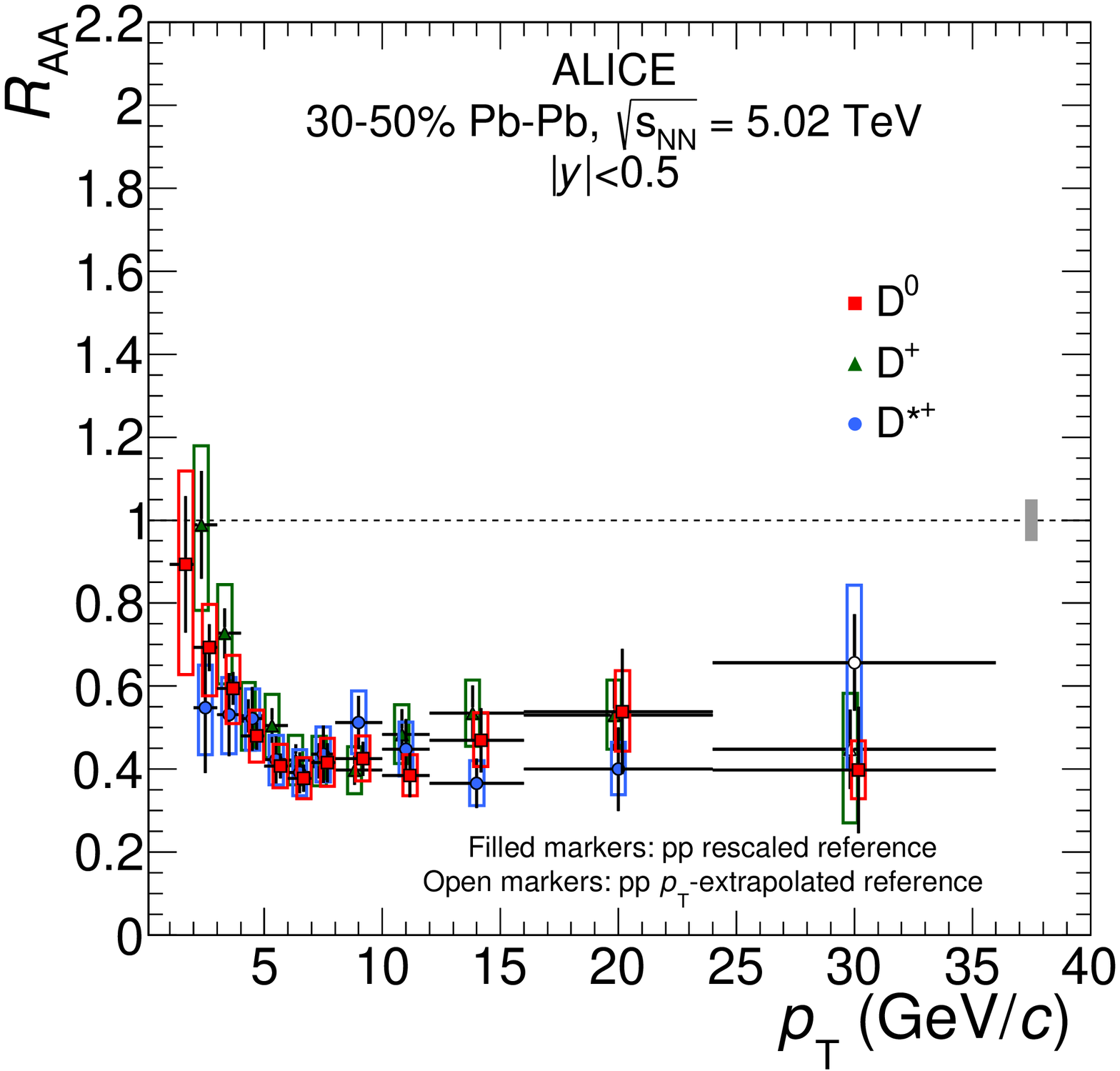}
\includegraphics[angle=0, width=0.43\textwidth]{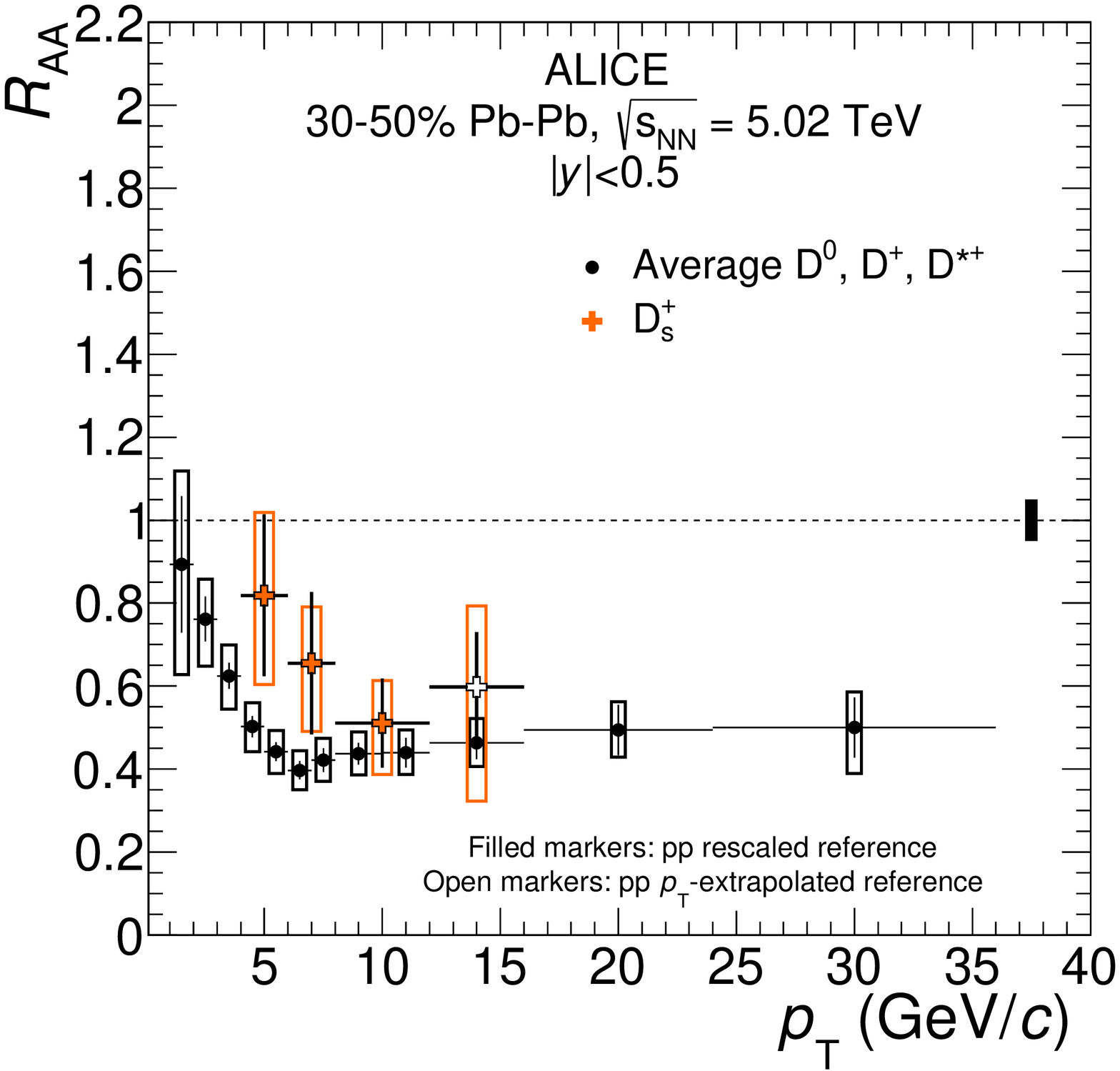}
\includegraphics[angle=0, width=0.43\textwidth]{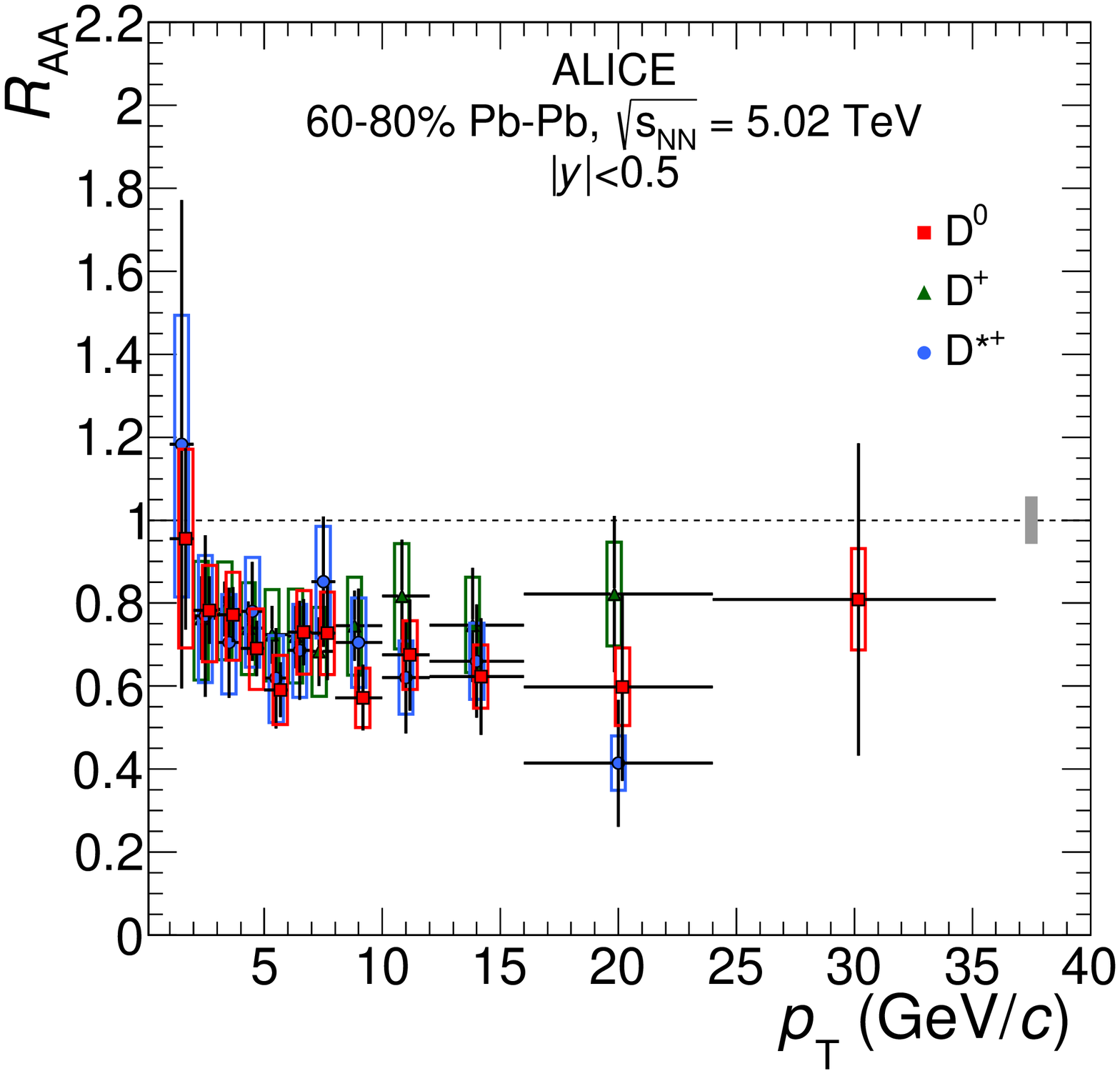}
\includegraphics[angle=0, width=0.43\textwidth]{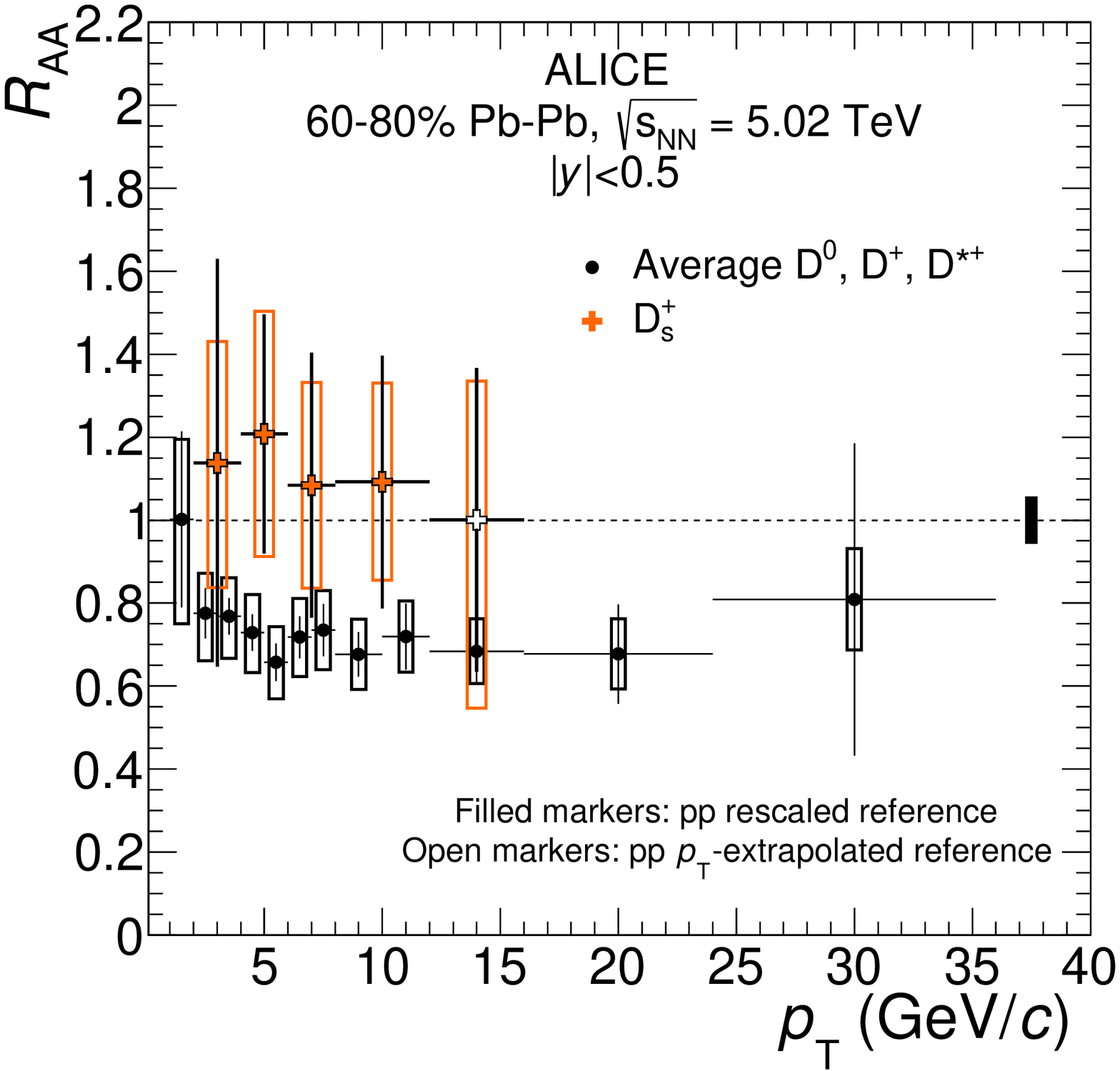}
 \end{center}
 \caption{$\RAA$ 
  of prompt $\Dzero$, $\Dplus$ and $\Dstar$ mesons (left-hand panels) and of prompt $\Ds$ mesons 
  compared with the average $\RAA$ of the non-strange D-meson states available in each $\pt$ interval (right-hand panels) for the 
0--10\%, 30--50\% and 60--80\% centrality classes. 
Statistical (bars),  systematic (empty boxes), and normalisation (shaded box around unity) 
uncertainties are shown. Filled markers are obtained with the pp rescaled reference, empty markers with the $\pt$-rescaled reference.}
 \label{DmesRaa} 
\end{figure} 

The $\Raa$ of prompt $\Dzero$, $\Dplus$ and $\Dstar$ mesons
is shown in the left-hand panels of Fig.~\ref{DmesRaa}, from central (top) to peripheral (bottom) collisions. The nuclear modification factors of the three D-meson species are compatible 
within statistical uncertainties, which are obtained by propagating those on the Pb--Pb yields and those of the pp reference. 
Their average was computed using the inverse of the quadratic sum of the relative statistical and uncorrelated systematic uncertainties as weights, in the $\pt$ intervals where more than one D-meson species is available, (Fig.~\ref{DmesRaa}, right-hand panels). 
The systematic uncertainties were propagated through the averaging procedure,
considering the contributions from the tracking efficiency, the beauty-hadron feed-down subtraction 
and the FONLL-based $\sqrts$-scaling of the pp cross section from
$\sqrts = 7~\TeV$  to 
$\sqrts = 5.02~\TeV$ as fully correlated uncertainties among the three D-meson species. 
The average nuclear modification factors in the 0--10\% and 30--50\% centrality classes (top and middle right-hand panels of Fig.~\ref{DmesRaa}) show a suppression that is
maximal at $\pt=6$--$10~\gev/c$, where a reduction of the yields by
a factor of about 5 and 2.5 with respect to the binary-scaled pp reference is observed in the two centrality classes, respectively.
The suppression gets smaller with decreasing $\pt$ for $\pt<6~\GeV/c$, and 
$\RAA$ is compatible with unity  in the interval $1<\pt<3~\gev/c$.
The average $\RAA$ in the 60--80\% centrality class shows a suppression by about 20--30\%, without a pronounced dependence on $\pt$.

\begin{figure}[!t]
 \begin{center}
\includegraphics[angle=0, width=0.49\textwidth]{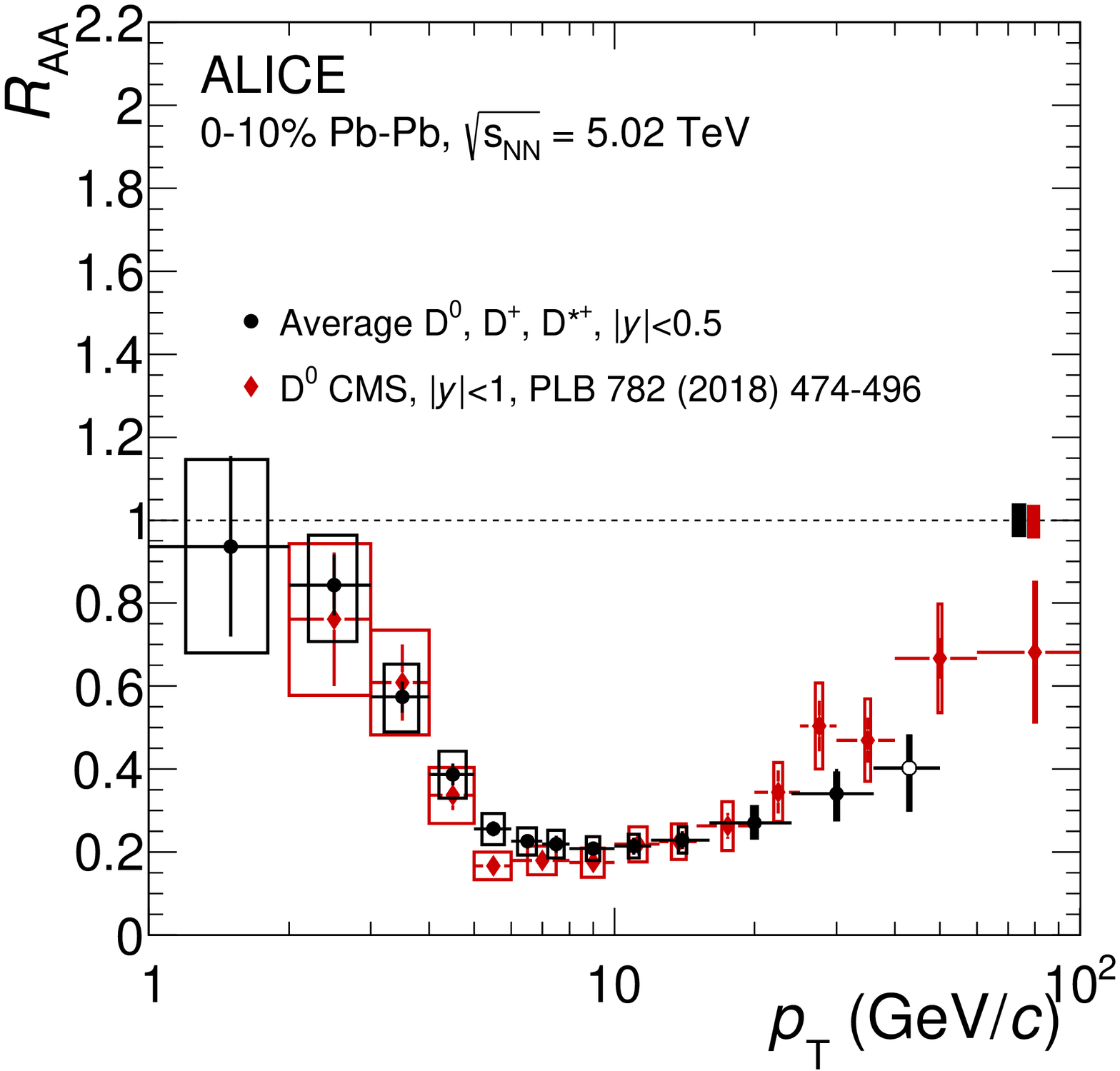}
\includegraphics[angle=0, width=0.49\textwidth]{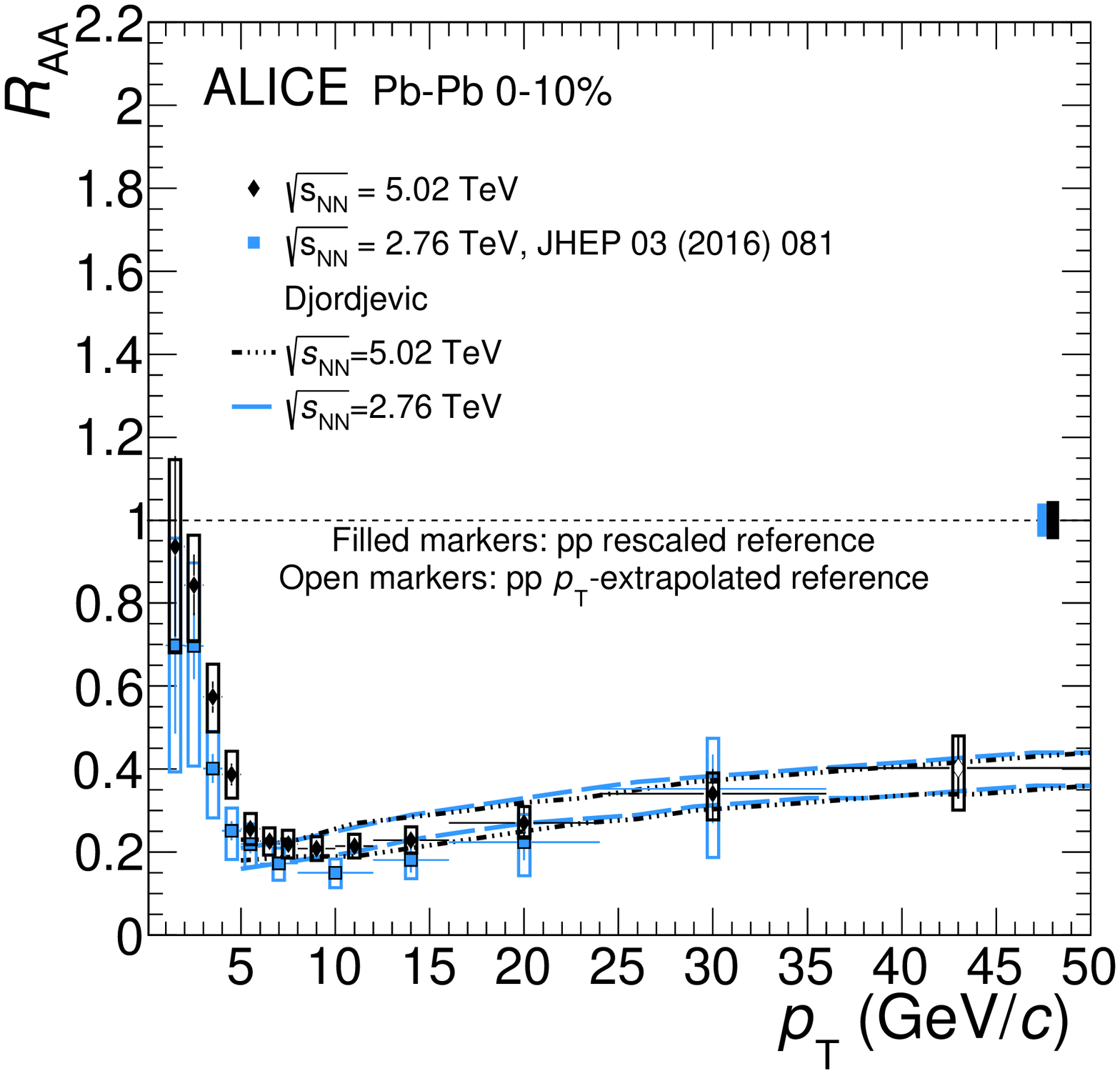}
 \end{center}
 \caption{Left panel:  average $\RAA$ of prompt $\Dzero$, $\Dplus$ and $\Dstar$ mesons by ALICE compared to $\RAA$ of prompt $\Dzero$ mesons by CMS~\cite{Sirunyan:2017xss} in the 0--10\% centrality class and at $\sqrtsNN= 5.02~\tev$. Statistical (bars), systematic (empty boxes), and normalisation (shaded box around unity) uncertainties are shown. Right panel: average $\RAA$ of $\Dzero$, $\Dplus$ and $\Dstar$ mesons compared with the Djordjevic model~\cite{Djordjevic:2015hra} in the 0-10\% centrality class at two collision energies. Statistical (bars), systematic (empty boxes), and normalisation (shaded box) uncertainties are shown. }
 \label{fig:DmesRaaVSCMSandEnergy} 
\end{figure}

The $\Raa$ of prompt $\Ds$ mesons is shown in the right-hand panels of Fig.~\ref{DmesRaa},
where it is compared with the average $\Raa$ of non-strange D mesons: the values are
larger for $\Ds$ mesons, but the two measurements are 
compatible within one standard deviation of the combined uncertainties, as is the case for the ratios shown in Fig.~\ref{DmesRatio}. 
The average $\RAA$ of prompt $\Dzero$, $\Dplus$ and $\Dstar$ in the 10\% most central collisions is compared with a measurement of prompt $\Dzero$ mesons by the CMS Collaboration~\cite{Sirunyan:2017xss} in the rapidity interval $|y|<1$ in Fig.~\ref{fig:DmesRaaVSCMSandEnergy} (left panel): the measurements are compatible in the common $\pt$ interval 2--50~$\gev/c$.
In the right panel of Fig.~\ref{fig:DmesRaaVSCMSandEnergy}, the nuclear modification factor of D mesons at $\sqrtsNN = 5.02~\tev$ in the 0--10\% centrality class is compared with the same measurement at 
$\sqrtsNN = 2.76~\tev$~\cite{Adam:2015sza}\footnote{The $T_{\rm AA}$  used to compute the D-meson $\Raa$ at $\sqrtsNN = 2.76~\tev$ in the 0--10\% centrality class and its uncertainty were updated with respect to~\cite{Adam:2015sza} according to the values reported in Ref.~\cite{ALICE-PUBLIC-2018-011}}.  
The measurement at $\sqrtsNN = 5.02~\tev$ have total uncertainties reduced 
by a factor of about two and extended $\pt$ coverage from 36 to 50~$\gev/c$. The suppression is compatible within uncertainties at the two energies, as also observed for charged particles~\cite{Acharya:2018qsh}.
 
The close similarity of the $\RAA$ measurements at the two energies was predicted by the Djordjevic model~\cite{Djordjevic:2015hra} (Fig.~\ref{fig:DmesRaaVSCMSandEnergy}, right panel), and it results from the combination of a higher medium temperature at 5.02~TeV (estimated to be about 7\% higher than at 2.76~TeV), which would decrease the $\RAA$ by about 10\%, with a harder $\pt$ distribution of charm quarks at 5.02~TeV, which would increase the $\RAA$ by about 5\% if the medium temperature were the same as at 2.76~TeV.

\begin{figure}[!t]
 \begin{center}
\includegraphics[angle=0, width=1\textwidth]{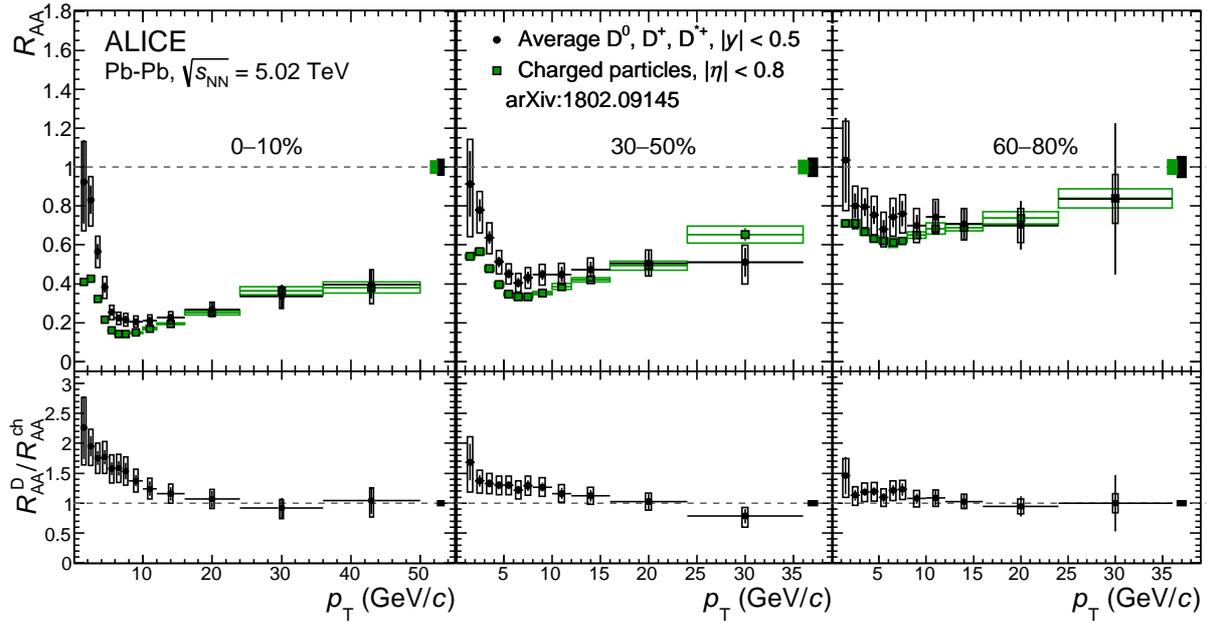}
 \end{center}
 \caption{Average $\RAA$ 
  of prompt $\Dzero$, $\Dplus$ and $\Dstar$ mesons in the 0--10\% (left), 30--50\% (middle) and 60--80\% (right) centrality classes at $\sqrtsNN=5.02~\tev$ compared to the $\RAA$ of charged particles in the same centrality classes~\cite{Acharya:2018qsh}. The ratios of the $\RAA$ are shown in the bottom panels. Statistical (bars), systematic (empty boxes), and normalisation (shaded box around unity) uncertainties are shown.}
 \label{DmesRaaVSPions} 
\end{figure}

As explained in Section~\ref{sec:intro}, the measurement of the $\RAA$ of open-charm mesons is essential to understand in-medium parton 
energy loss, in particular its colour-charge and quark-mass dependence. In Fig.~\ref{DmesRaaVSPions},
 the $\RAA$ of prompt D mesons is compared with that of charged particles in the same $\pt$ intervals, at the same energy and in 
 the same centrality classes~\cite{Acharya:2018qsh}. The ratio of their nuclear modification factors is displayed 
 in the bottom panels, for the three centrality classes. The $\RAA$ of D mesons and charged particles differ by more than $2\,\sigma$ of the combined statistical and systematic uncertainties
 in all the $\pt$ intervals within $3<\pt<8~\gev/c$ in central collisions.
 The difference is less than $2\,\sigma$ in this range for semi-central collisions, 
 while the two $\RAA$ are the same within $1\,\sigma$ for $\pt>10~\gev/c$ in both central and semi-central collisions.
In the 60--80\% class the measurements are compatible in the common $\pt$ interval.
The interpretation of the difference observed for $\pt < 8~\gev/c$ in central and semi-central 
collisions is not straightforward, because several factors can play a role in defining the shape 
of the $\RAA$. 

In presence of a colour-charge and quark-mass dependent energy loss, 
the harder $\pt$ distribution and the harder fragmentation function of charm quarks compared 
to those of light quarks and gluons should lead to similar values of D-meson and pion $\RAA$, 
as discussed in~\cite{Djordjevic:2013pba}. Since the pions are the dominant contribution in the inclusive charged-particle yields, this statement is expected to be still valid for the 
comparison of the D-meson and the charged particle $\Raa$. In addition, it should be considered that the yield of light-flavour hadrons 
could have a substantial contribution up to transverse momenta of 
about 2--3~$\gev/c$ from soft production processes, such as the break-down of participant nucleons into quarks and gluons that subsequently hadronise. 
This component scales with the number of participants rather than the number of binary collisions. 
Finally, the effects of radial flow and hadronisation via recombination, as well as initial-state effects, 
could affect D-meson and light-hadron yields differently at a given $\pt$. 

\begin{figure}[!h]
 \begin{center}
\includegraphics[angle=0, width=0.43\textwidth]{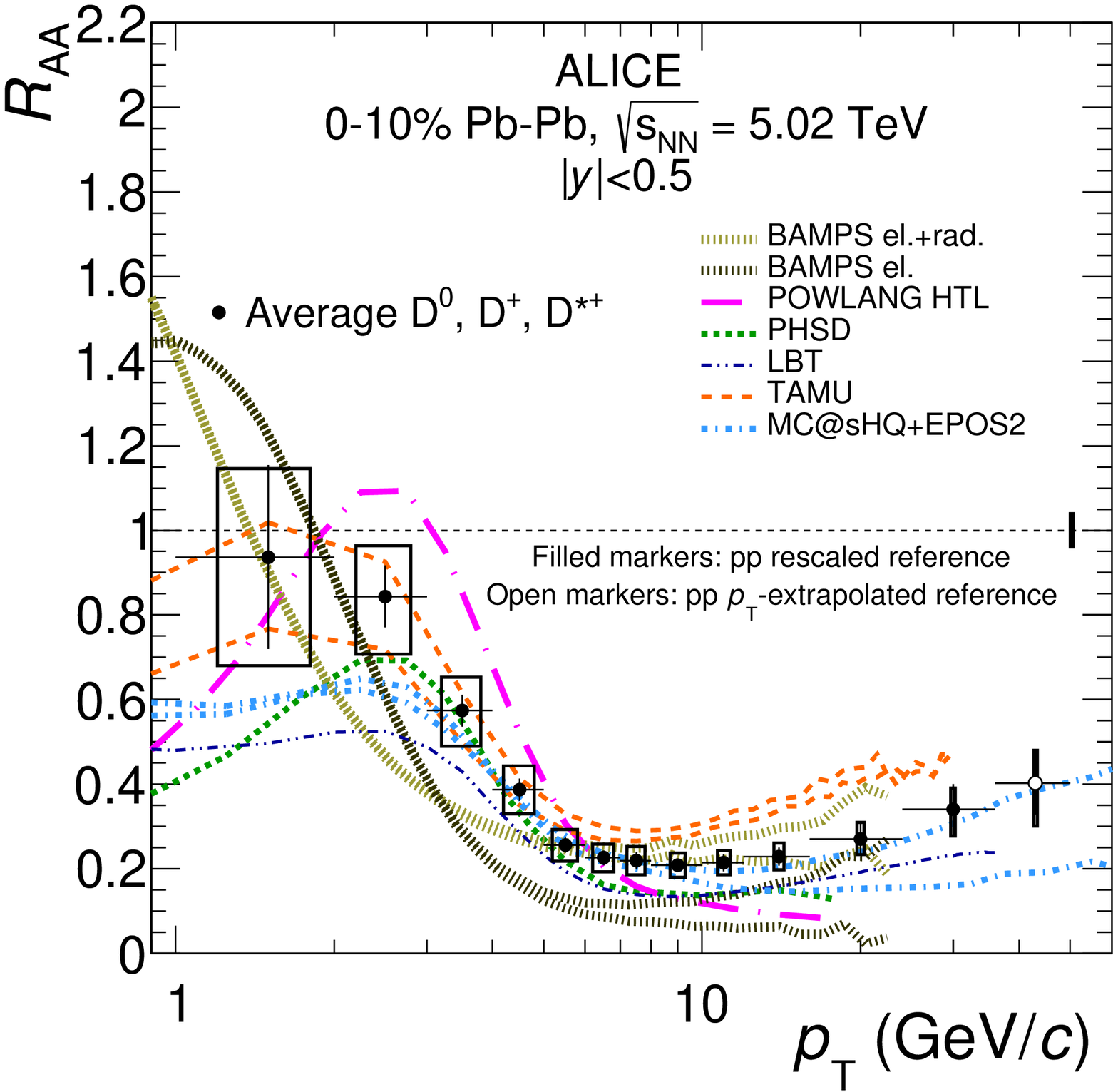}
\includegraphics[angle=0, width=0.43\textwidth]{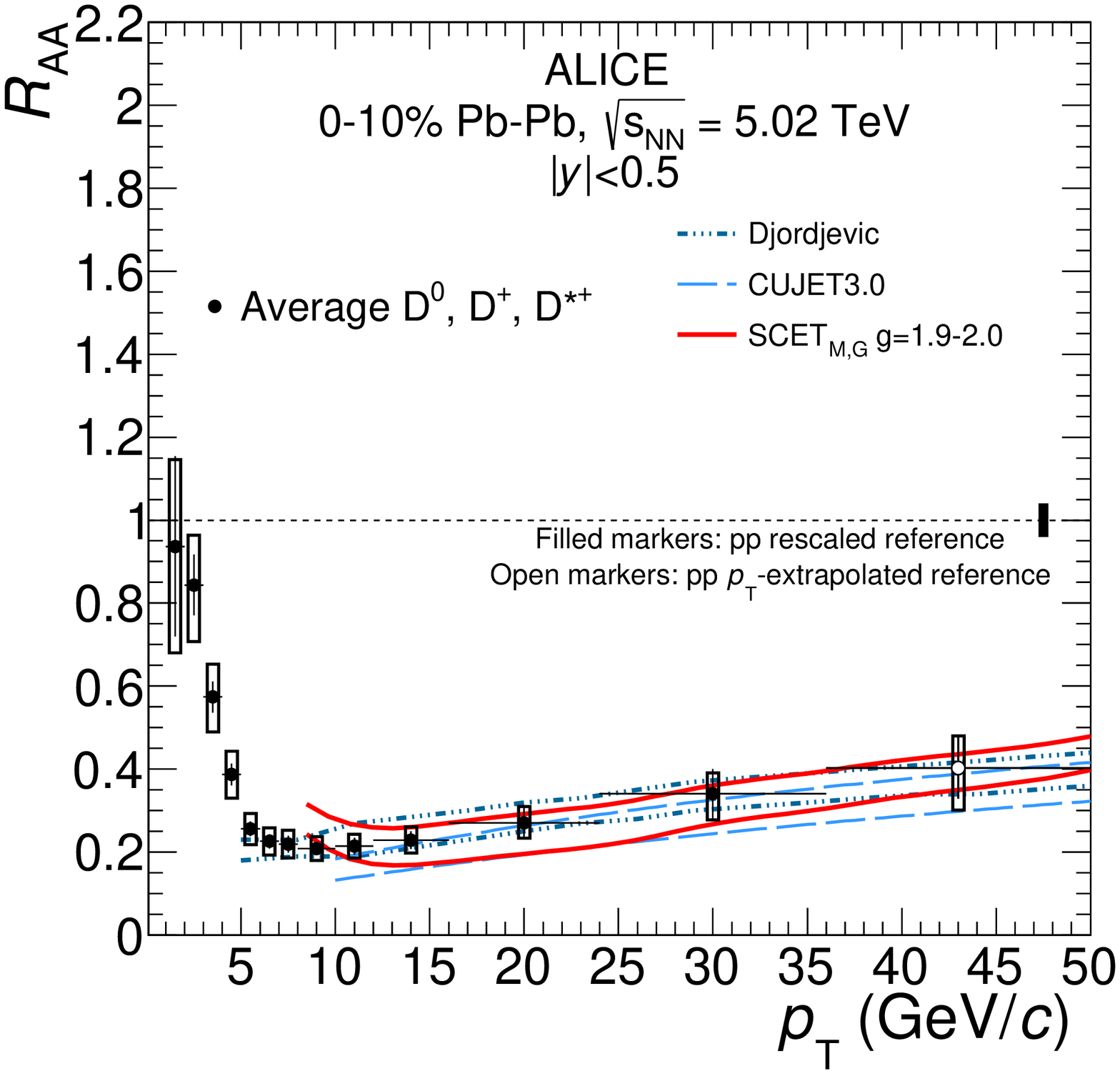}
\includegraphics[angle=0, width=0.43\textwidth]{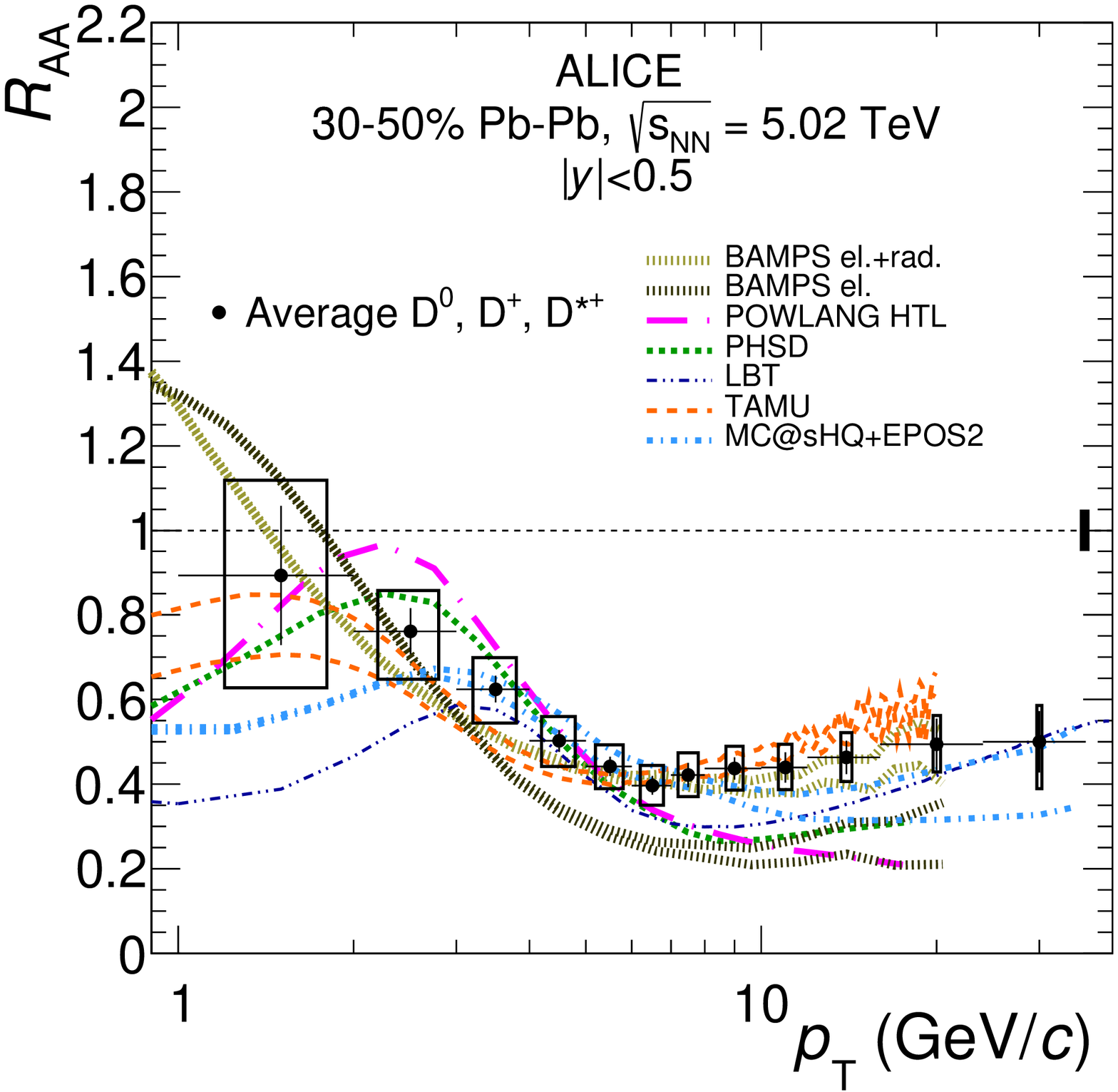}
\includegraphics[angle=0, width=0.43\textwidth]{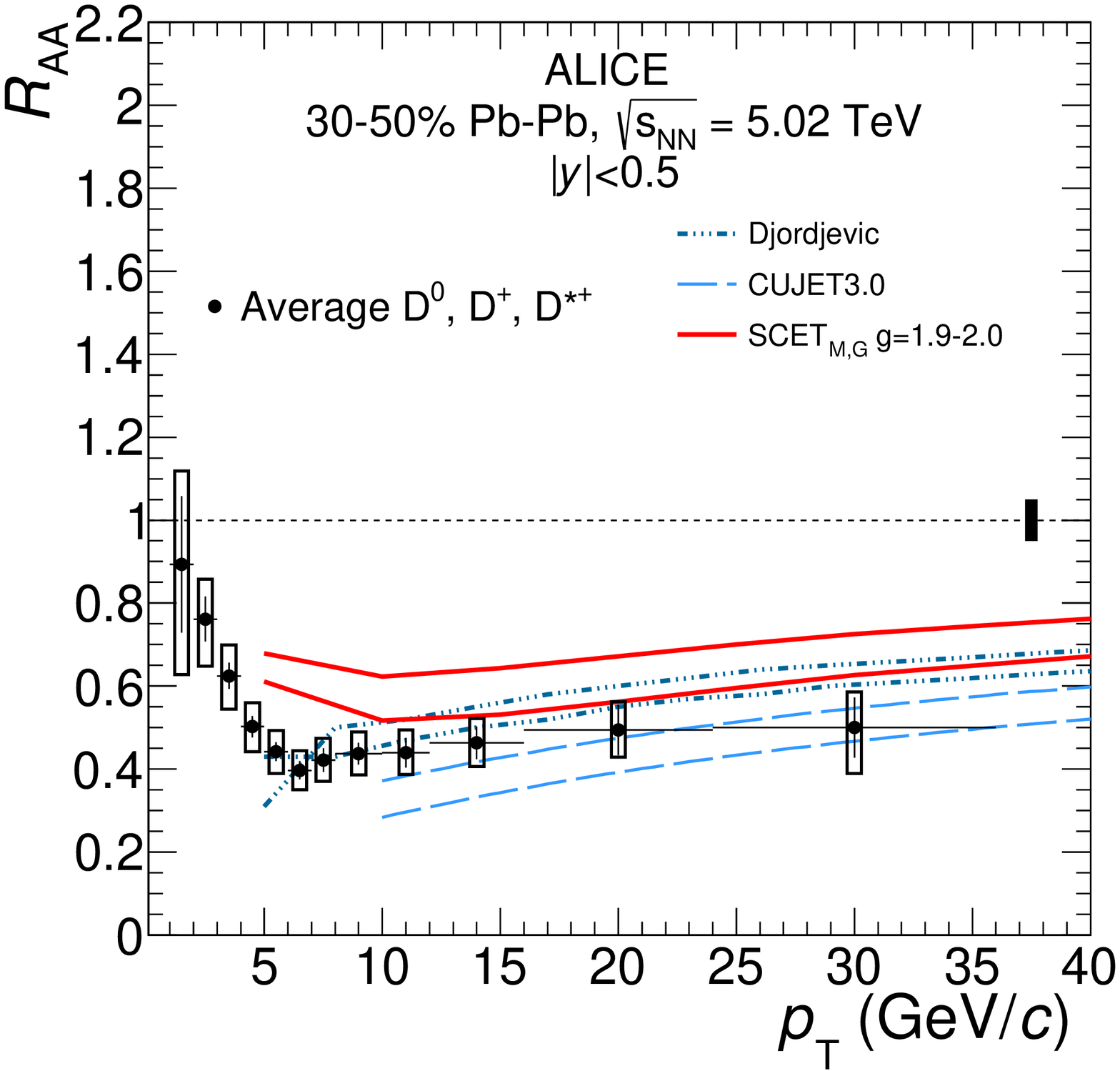}
\includegraphics[angle=0, width=0.43\textwidth]{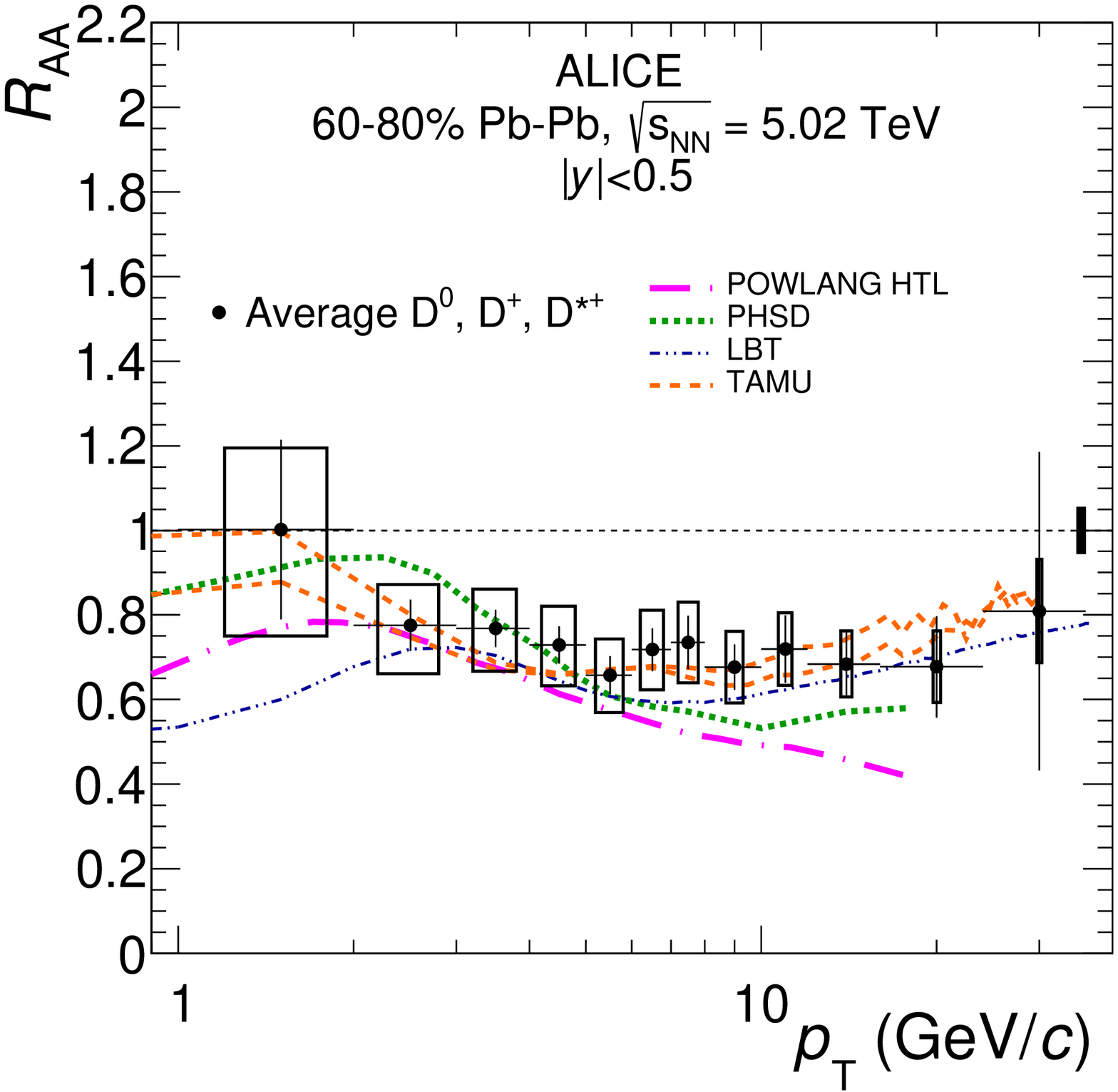}
\includegraphics[angle=0, width=0.43\textwidth]{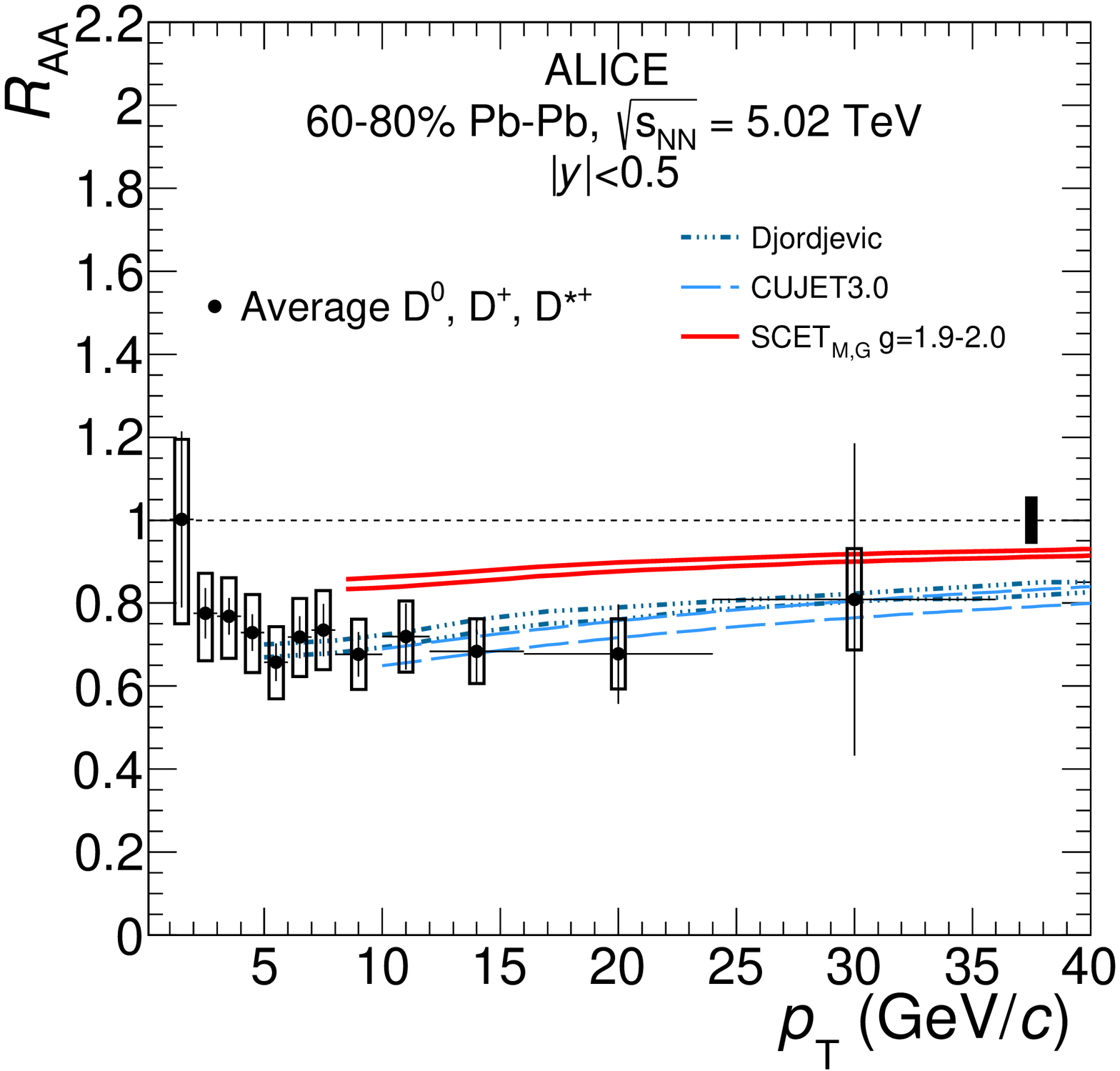}
 \end{center}
 \caption{Average $\RAA$ of $\Dzero$, $\Dplus$ and $\Dstar$ mesons compared with model calculations. 
 The three rows refer to the 0--10\%, 30--50\% and 60--80\% centrality classes. 
 The left panels show models based on heavy-quark transport, while the right panels show models based on pQCD energy loss. 
 Model nomenclature and references:
BAMPS~\cite{Uphoff:2014hza},  
CUJET3.0~\cite{Xu:2015bbz},
Djordjevic~\cite{Djordjevic:2015hra},
LBT~\cite{Cao:2017hhk},
MC@sHQ+EPOS2~\cite{Nahrgang:2013xaa},
PHSD~\cite{Song:2015ykw} 
POWLANG~\cite{Beraudo:2014boa}, 
SCET~\cite{Kang:2016ofv},
TAMU~\cite{He:2014cla}.
Some of the models are presented with two lines with the same style and colour, which encompass the model uncertainty band.}
 \label{DmesRaaWithModels} 
\end{figure}

The average $\RAA$ of the three non-strange D-meson species in the three centrality classes are compared with theoretical models in Fig.~\ref{DmesRaaWithModels}. 
Models based on heavy-quark transport and models based on perturbative QCD calculations of high-$\pt$ parton energy loss are shown in the left and in the right panels, respectively.
Transport models in the left panels include: BAMPS el.~\cite{Uphoff:2014hza}, POWLANG~\cite{Beraudo:2014boa} and TAMU~\cite{He:2014cla}, in which the interactions are only described by collisional (i.e.\,elastic) processes; 
BAMPS el.+rad.~\cite{Uphoff:2014hza}, LBT~\cite{Cao:2017hhk}, MC@sHQ+EPOS2~\cite{Nahrgang:2013xaa} and PHSD~\cite{Song:2015ykw}, in which also energy loss from medium-induced gluon radiation
is considered, in addition to collisional process.
In the right panels, the CUJET3.0~\cite{Xu:2015bbz} and Djordjevic~\cite{Djordjevic:2015hra} models include both radiative and collisional energy loss processes, while
the SCET~\cite{Kang:2016ofv} model implements medium-induced gluon radiation via modified splitting functions with finite quark masses\footnote{The SCET curves reported here differ from 
those of Ref.~\cite{Kang:2016ofv} because the latter used an extrapolation of the charged-particle multiplicity at $\sqrtsNN=5.02~\tev$, while now the measured values are used.}.
All models, with the exception of BAMPS and CUJET3.0, include a nuclear modification of the parton distribution functions.
The LBT, MC@sHQ, PHSD, POWLANG and TAMU
models include a contribution of hadronisation via quark recombination, in addition to independent fragmentation. 
Most of the models provide a fair description of the data in the region $\pt<10~\gev/c$ in central collisions 
(except for BAMPS el., where the radiative term is missing), 
but many of them (LBT, PHSD, POWLANG and SCET) provide a worse description of non-central collisions.
In the high-$\pt$ region above $10~\gev/c$ only the BAMPS el.+rad., CUJET3.0, Djordjevic, MC@sHQ+EPOS2 and SCET models can describe the data in central collisions.
The CUJET3.0 and Djordjevic models provide a 
fair description of the $\RAA$ in all three centrality classes for $\pt > 10~\gev/c$, where radiative energy loss is expected to be the dominant interaction mechanism, suggesting that the dependence of radiative energy loss on the path length in the hot and dense medium is well understood. 

\begin{figure}[!t]
 \begin{center}
\includegraphics[angle=0, width=0.49\textwidth]{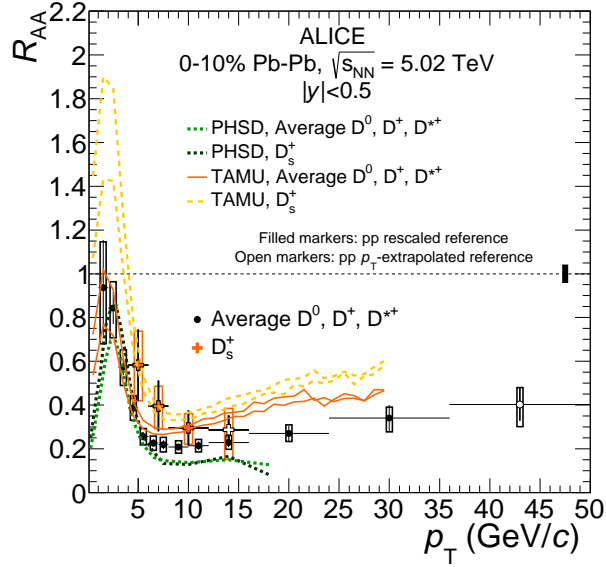}
 \end{center}
 \caption{Average $\RAA$ of $\Dzero$, $\Dplus$ and $\Dstar$ mesons and $\RAA$ of $\Ds$ mesons in the 0--10\% centrality class compared with the PHSD~\cite{Song:2015ykw}  and TAMU~\cite{He:2014cla} model calculations.}
 \label{DandDsRaaWithModels} 
\end{figure}

In Fig.~\ref{DandDsRaaWithModels},  the non-strange and strange D-meson $\RAA$ are compared with  
the models that provide both observables. An increase of the $\Ds$ $\RAA$ is expected in the two models, PHSD and TAMU, in particular for $\pt<5~\gev/c$, with respect to non-strange D mesons. This increase is induced by hadronisation via quark recombination in the QGP, as well as by different 
interaction cross sections for non-strange D and for $\Ds$ in the hadronic phase of the system evolution. 
In the transverse momentum interval covered by the $\Ds$ measurement ($\pt>4~\gev/c$),  the PHSD model predicts the effect to be very small, while the TAMU model predicts a sizeable difference of about 30\% up to about $8~\gev/c$,  similar to the trend shown by the data.
\clearpage 

\begin{figure}[!b]
 \begin{center}
\includegraphics[angle=0, width=0.49\textwidth]{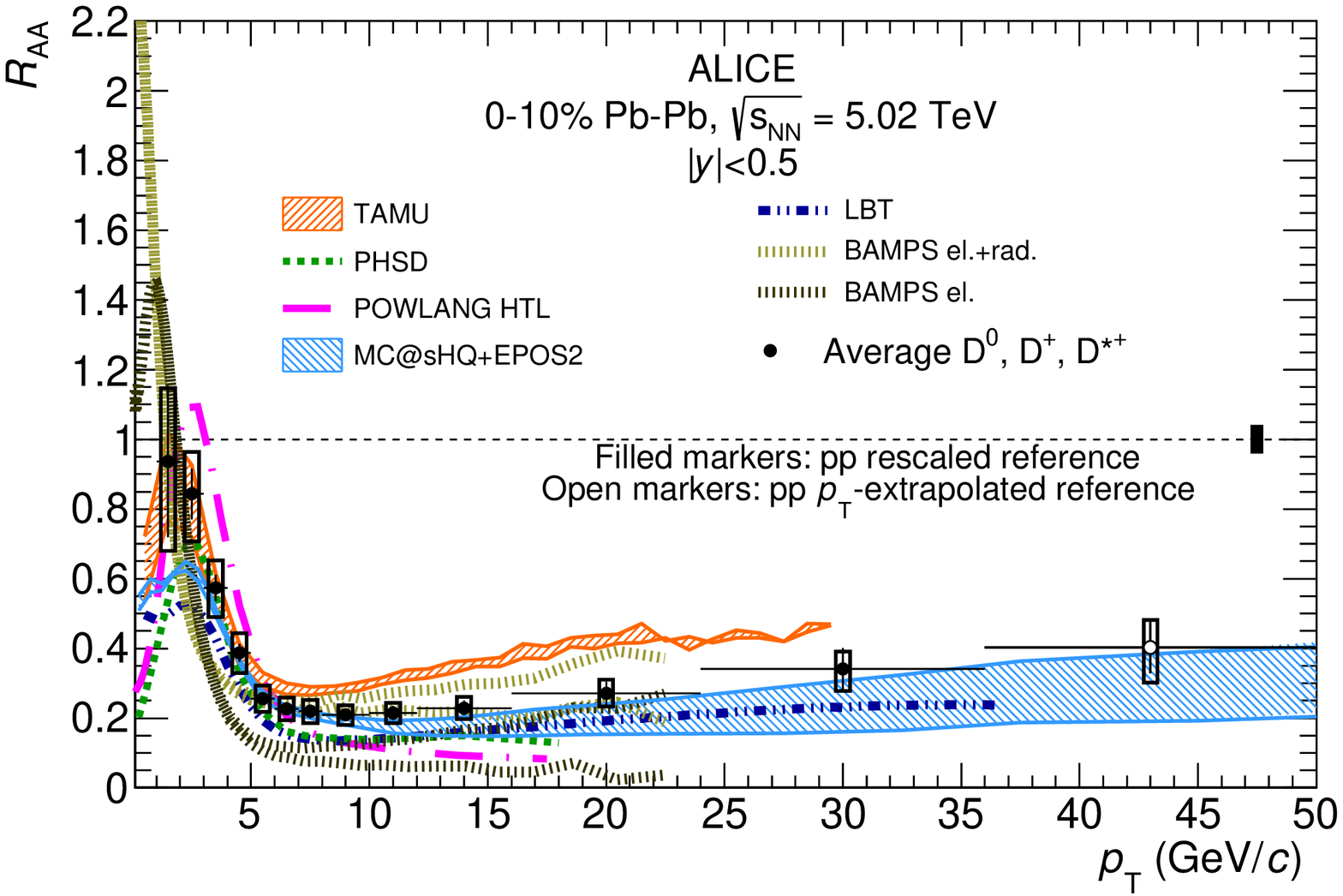}
\includegraphics[angle=0, width=0.49\textwidth]{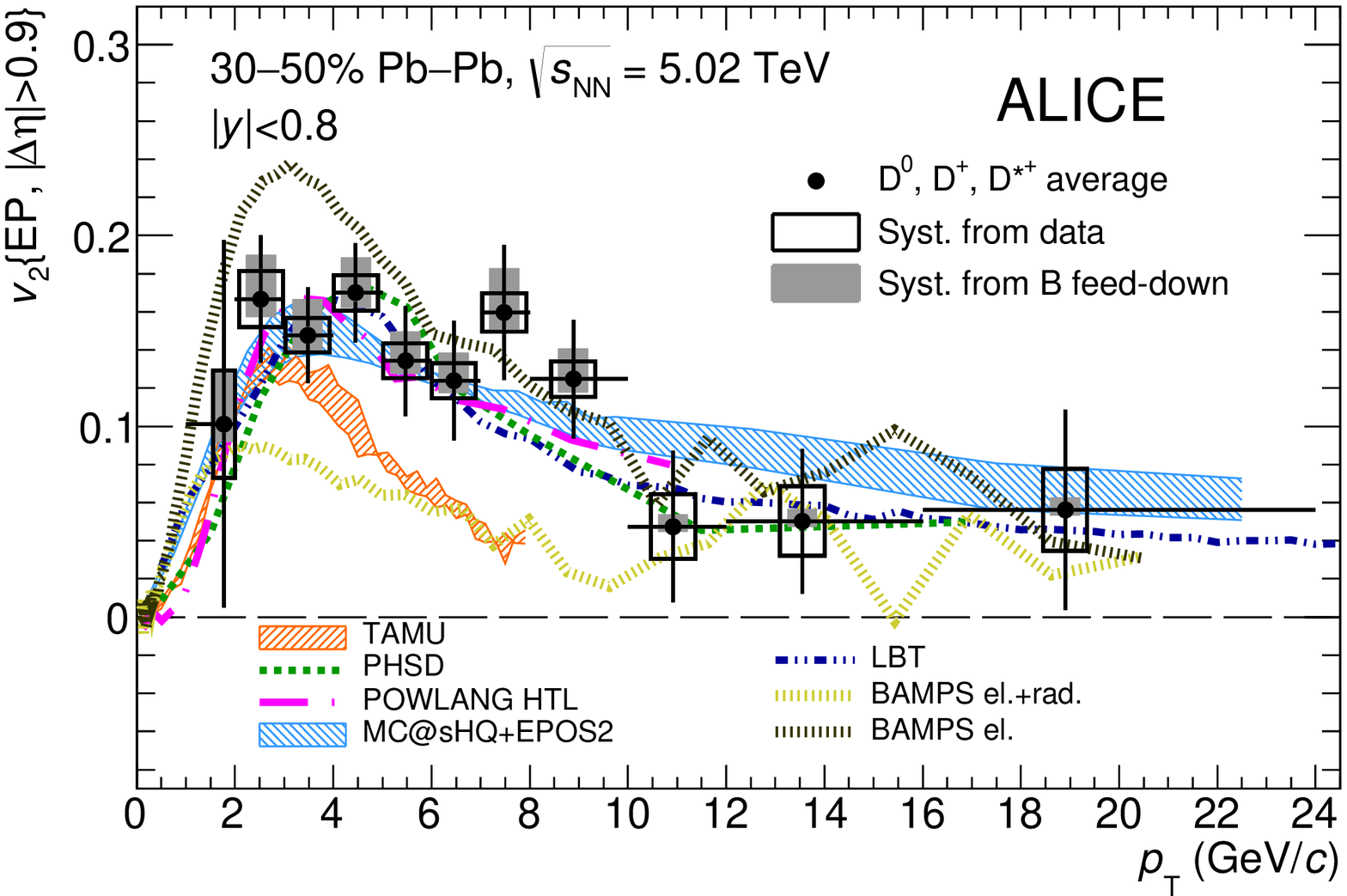}
 \end{center}
 \caption{Average $\RAA$ of $\Dzero$, $\Dplus$ and $\Dstar$ mesons in the 0--10\% centrality class (left) and their average elliptic flow $v_2$ in the 30--50\% centrality class (right)~\cite{Acharya:2017qps}, compared with models that have predictions for both observables at low $\pt$.}
 \label{RAAandv2} 
\end{figure}

The simultaneous comparison of $\RAA$ and elliptic flow $v_2$ measurements at $\sqrtsNN=5.02~\tev$~\cite{Acharya:2017qps} with models can provide more stringent constraints to the implementation of the interaction and hadronisation processes for heavy quarks in the QGP. 
The comparison with models that compute both observables is shown in Fig.~\ref{RAAandv2} for the $\RAA$  
and $v_2$, in the 0--10\% and 30--50\% centrality classes, respectively. The TAMU model overestimates $\RAA$ and underestimates $v_2$ at high $\pt$, probably because it does not include radiative energy loss.
The BAMPS el.\,model overestimates the maximum flow while underestimating the $\RAA$ value at high $\pt$.  The radiative energy loss contribution in BAMPS el.+rad.\,improves the description of $\RAA$ but gives $v_2$ values lower than the data. The LBT, PHSD, POWLANG and MC@sHQ models provide instead a fair description of $v_2$.
Nevertheless, energy loss is overestimated at high $\pt$ in the 0--10\% centrality classes (but also in semi-central events) by PHSD, POWLANG and LBT, while at low $\pt$ the measured $\RAA$ is slightly higher than what predicted within LBT, PHSD and MC@sHQ.

\section{Summary}
\label{sec:summary}

We have presented measurements of the $\pt$-differential production yields of
prompt $\Dzero$, $\Dplus$, $\Dstar$ and $\Ds$ mesons at central rapidity 
in Pb--Pb collisions in the three centrality classes 0--10\%, 30--50\% and 60--80\% at a centre-of-mass energy per nucleon pair $\sqrtsNN=5.02~\tev$.

The average $\RAA$ of the three non-strange D-meson species  
shows minimum values of 0.2 and 0.4 
in the centrality classes 0--10\% and 30--50\%, respectively, at $\pt$ of 6--$10~\gevc$. 
$\RAA$ increases for $\pt<6~\GeV/c$, and 
it is compatible with unity at $1<\pt<3~\gev/c$.
The average $\RAA$ values are compatible with those 
measured at $\sqrtsNN=2.76~\tev$ and they 
have smaller uncertainties 
by a factor of about two, as well as extended $\pt$ coverage up to 50~$\gev/c$ 
in central collisions.
The similarity of the $\RAA$ values at the two energies was predicted by the Djordjevic model, 
and it results from the combination of a higher medium temperature at 5.02~TeV (estimated to be about 7\% higher than at 2.76~TeV)
with a harder $\pt$ distribution of charm quarks at 5.02~TeV.

In central and semi-central collisions the average $\RAA$ of non-strange D mesons is compatible with that of charged particles for $\pt>6~\gev/c$, 
while it is larger at lower $\pt$. The $\Raa$ of $\Ds$ mesons have generally larger 
central values than those of the average of non-strange D mesons, but the 
two measurements are compatible within about one standard deviation of the combined uncertainties. 

The $\RAA$ of non-strange D mesons at high $\pt$ (above $10~\gev/c$) is fairly described in the three centrality classes 
by model calculations that include both radiative and collisional energy loss.  This indicates that the centrality dependence 
of radiative energy loss, which is the dominant contribution at high $\pt$, is under good theoretical control.
The $\RAA$  in the transverse momentum region below $10~\gev/c$ 
is described by several  transport model 
calculations  in central collisions, but most models fail in describing the centrality 
dependence of $\RAA$ and in describing simultaneously $\RAA$ and the elliptic flow coefficient $v_2$.
Therefore, the measurements provide significant constraints for the understanding of the interaction of charm quarks with the high-density QCD medium, especially at low and intermediate $\pt$, where the $\RAA$ is the result of a more complex interplay among several effects.

\newenvironment{acknowledgement}{\relax}{\relax}
\begin{acknowledgement}
\section*{Acknowledgements}

The ALICE Collaboration would like to thank all its engineers and technicians for their invaluable contributions to the construction of the experiment and the CERN accelerator teams for the outstanding performance of the LHC complex.
The ALICE Collaboration gratefully acknowledges the resources and support provided by all Grid centres and the Worldwide LHC Computing Grid (WLCG) collaboration.
The ALICE Collaboration acknowledges the following funding agencies for their support in building and running the ALICE detector:
A. I. Alikhanyan National Science Laboratory (Yerevan Physics Institute) Foundation (ANSL), State Committee of Science and World Federation of Scientists (WFS), Armenia;
Austrian Academy of Sciences and Nationalstiftung f\"{u}r Forschung, Technologie und Entwicklung, Austria;
Ministry of Communications and High Technologies, National Nuclear Research Center, Azerbaijan;
Conselho Nacional de Desenvolvimento Cient\'{\i}fico e Tecnol\'{o}gico (CNPq), Universidade Federal do Rio Grande do Sul (UFRGS), Financiadora de Estudos e Projetos (Finep) and Funda\c{c}\~{a}o de Amparo \`{a} Pesquisa do Estado de S\~{a}o Paulo (FAPESP), Brazil;
Ministry of Science \& Technology of China (MSTC), National Natural Science Foundation of China (NSFC) and Ministry of Education of China (MOEC) , China;
Ministry of Science and Education, Croatia;
Ministry of Education, Youth and Sports of the Czech Republic, Czech Republic;
The Danish Council for Independent Research | Natural Sciences, the Carlsberg Foundation and Danish National Research Foundation (DNRF), Denmark;
Helsinki Institute of Physics (HIP), Finland;
Commissariat \`{a} l'Energie Atomique (CEA) and Institut National de Physique Nucl\'{e}aire et de Physique des Particules (IN2P3) and Centre National de la Recherche Scientifique (CNRS), France;
Bundesministerium f\"{u}r Bildung, Wissenschaft, Forschung und Technologie (BMBF) and GSI Helmholtzzentrum f\"{u}r Schwerionenforschung GmbH, Germany;
General Secretariat for Research and Technology, Ministry of Education, Research and Religions, Greece;
National Research, Development and Innovation Office, Hungary;
Department of Atomic Energy Government of India (DAE), Department of Science and Technology, Government of India (DST), University Grants Commission, Government of India (UGC) and Council of Scientific and Industrial Research (CSIR), India;
Indonesian Institute of Science, Indonesia;
Centro Fermi - Museo Storico della Fisica e Centro Studi e Ricerche Enrico Fermi and Istituto Nazionale di Fisica Nucleare (INFN), Italy;
Institute for Innovative Science and Technology , Nagasaki Institute of Applied Science (IIST), Japan Society for the Promotion of Science (JSPS) KAKENHI and Japanese Ministry of Education, Culture, Sports, Science and Technology (MEXT), Japan;
Consejo Nacional de Ciencia (CONACYT) y Tecnolog\'{i}a, through Fondo de Cooperaci\'{o}n Internacional en Ciencia y Tecnolog\'{i}a (FONCICYT) and Direcci\'{o}n General de Asuntos del Personal Academico (DGAPA), Mexico;
Nederlandse Organisatie voor Wetenschappelijk Onderzoek (NWO), Netherlands;
The Research Council of Norway, Norway;
Commission on Science and Technology for Sustainable Development in the South (COMSATS), Pakistan;
Pontificia Universidad Cat\'{o}lica del Per\'{u}, Peru;
Ministry of Science and Higher Education and National Science Centre, Poland;
Korea Institute of Science and Technology Information and National Research Foundation of Korea (NRF), Republic of Korea;
Ministry of Education and Scientific Research, Institute of Atomic Physics and Romanian National Agency for Science, Technology and Innovation, Romania;
Joint Institute for Nuclear Research (JINR), Ministry of Education and Science of the Russian Federation and National Research Centre Kurchatov Institute, Russia;
Ministry of Education, Science, Research and Sport of the Slovak Republic, Slovakia;
National Research Foundation of South Africa, South Africa;
Centro de Aplicaciones Tecnol\'{o}gicas y Desarrollo Nuclear (CEADEN), Cubaenerg\'{\i}a, Cuba and Centro de Investigaciones Energ\'{e}ticas, Medioambientales y Tecnol\'{o}gicas (CIEMAT), Spain;
Swedish Research Council (VR) and Knut \& Alice Wallenberg Foundation (KAW), Sweden;
European Organization for Nuclear Research, Switzerland;
National Science and Technology Development Agency (NSDTA), Suranaree University of Technology (SUT) and Office of the Higher Education Commission under NRU project of Thailand, Thailand;
Turkish Atomic Energy Agency (TAEK), Turkey;
National Academy of  Sciences of Ukraine, Ukraine;
Science and Technology Facilities Council (STFC), United Kingdom;
National Science Foundation of the United States of America (NSF) and United States Department of Energy, Office of Nuclear Physics (DOE NP), United States of America.
\end{acknowledgement}


\bibliographystyle{utphys}
\bibliography{DRAA2015Paper.bib}

\newpage
\appendix
\section{The ALICE Collaboration}
\label{app:collab}

\begingroup
\small
\begin{flushleft}
S.~Acharya\Irefn{org138}\And 
F.T.-.~Acosta\Irefn{org22}\And 
D.~Adamov\'{a}\Irefn{org93}\And 
J.~Adolfsson\Irefn{org80}\And 
M.M.~Aggarwal\Irefn{org97}\And 
G.~Aglieri Rinella\Irefn{org36}\And 
M.~Agnello\Irefn{org33}\And 
N.~Agrawal\Irefn{org48}\And 
Z.~Ahammed\Irefn{org138}\And 
S.U.~Ahn\Irefn{org76}\And 
S.~Aiola\Irefn{org143}
% ...
\And
\renewcommand\labelenumi{\textsuperscript{\theenumi}~}

\section*{Affiliation notes}
\renewcommand\theenumi{\roman{enumi}}
\begin{Authlist}
\item \Adef{orgI}Dipartimento DET del Politecnico di Torino, Turin, Italy
\item \Adef{orgII}M.V. Lomonosov Moscow State University, D.V. Skobeltsyn Institute of Nuclear, Physics, Moscow, Russia
\item \Adef{orgIII}Department of Applied Physics, Aligarh Muslim University, Aligarh, India
\item \Adef{orgIV}Institute of Theoretical Physics, University of Wroclaw, Poland
\end{Authlist}

\section*{Collaboration Institutes}
\renewcommand\theenumi{\arabic{enumi}~}
\begin{Authlist}
\item \Idef{org1}A.I. Alikhanyan National Science Laboratory (Yerevan Physics Institute) Foundation, Yerevan, Armenia
\item \Idef{org2}Benem\'{e}rita Universidad Aut\'{o}noma de Puebla, Puebla, Mexico
\item \Idef{org3}Bogolyubov Institute for Theoretical Physics, National Academy of Sciences of Ukraine, Kiev, Ukraine
\item \Idef{org4}Bose Institute, Department of Physics  and Centre for Astroparticle Physics and Space Science (CAPSS), Kolkata, India
\item \Idef{org5}Budker Institute for Nuclear Physics, Novosibirsk, Russia
\item \Idef{org6}California Polytechnic State University, San Luis Obispo, California, United States
\item \Idef{org7}Central China Normal University, Wuhan, China
\item \Idef{org8}Centre de Calcul de l'IN2P3, Villeurbanne, Lyon, France
\item \Idef{org9}Centro de Aplicaciones Tecnol\'{o}gicas y Desarrollo Nuclear (CEADEN), Havana, Cuba
\item \Idef{org10}Centro de Investigaci\'{o}n y de Estudios Avanzados (CINVESTAV), Mexico City and M\'{e}rida, Mexico
\item \Idef{org11}Centro Fermi - Museo Storico della Fisica e Centro Studi e Ricerche ``Enrico Fermi', Rome, Italy
\item \Idef{org12}Chicago State University, Chicago, Illinois, United States
\item \Idef{org13}China Institute of Atomic Energy, Beijing, China
\item \Idef{org14}Chonbuk National University, Jeonju, Republic of Korea
\item \Idef{org15}Comenius University Bratislava, Faculty of Mathematics, Physics and Informatics, Bratislava, Slovakia
\item \Idef{org16}COMSATS Institute of Information Technology (CIIT), Islamabad, Pakistan
\item \Idef{org17}Creighton University, Omaha, Nebraska, United States
\item \Idef{org18}Department of Physics, Aligarh Muslim University, Aligarh, India
\item \Idef{org19}Department of Physics, Ohio State University, Columbus, Ohio, United States
\item \Idef{org20}Department of Physics, Pusan National University, Pusan, Republic of Korea
\item \Idef{org21}Department of Physics, Sejong University, Seoul, Republic of Korea
\item \Idef{org22}Department of Physics, University of California, Berkeley, California, United States
\item \Idef{org23}Department of Physics, University of Oslo, Oslo, Norway
\item \Idef{org24}Department of Physics and Technology, University of Bergen, Bergen, Norway
\item \Idef{org25}Dipartimento di Fisica dell'Universit\`{a} 'La Sapienza' and Sezione INFN, Rome, Italy
\item \Idef{org26}Dipartimento di Fisica dell'Universit\`{a} and Sezione INFN, Cagliari, Italy
\item \Idef{org27}Dipartimento di Fisica dell'Universit\`{a} and Sezione INFN, Trieste, Italy
\item \Idef{org28}Dipartimento di Fisica dell'Universit\`{a} and Sezione INFN, Turin, Italy
\item \Idef{org29}Dipartimento di Fisica e Astronomia dell'Universit\`{a} and Sezione INFN, Bologna, Italy
\item \Idef{org30}Dipartimento di Fisica e Astronomia dell'Universit\`{a} and Sezione INFN, Catania, Italy
\item \Idef{org31}Dipartimento di Fisica e Astronomia dell'Universit\`{a} and Sezione INFN, Padova, Italy
\item \Idef{org32}Dipartimento di Fisica `E.R.~Caianiello' dell'Universit\`{a} and Gruppo Collegato INFN, Salerno, Italy
\item \Idef{org33}Dipartimento DISAT del Politecnico and Sezione INFN, Turin, Italy
\item \Idef{org34}Dipartimento di Scienze e Innovazione Tecnologica dell'Universit\`{a} del Piemonte Orientale and INFN Sezione di Torino, Alessandria, Italy
\item \Idef{org35}Dipartimento Interateneo di Fisica `M.~Merlin' and Sezione INFN, Bari, Italy
\item \Idef{org36}European Organization for Nuclear Research (CERN), Geneva, Switzerland
\item \Idef{org37}Faculty of Engineering and Science, Western Norway University of Applied Sciences, Bergen, Norway
\item \Idef{org38}Faculty of Nuclear Sciences and Physical Engineering, Czech Technical University in Prague, Prague, Czech Republic
\item \Idef{org39}Faculty of Science, P.J.~\v{S}af\'{a}rik University, Ko\v{s}ice, Slovakia
\item \Idef{org40}Frankfurt Institute for Advanced Studies, Johann Wolfgang Goethe-Universit\"{a}t Frankfurt, Frankfurt, Germany
\item \Idef{org41}Gangneung-Wonju National University, Gangneung, Republic of Korea
\item \Idef{org42}Gauhati University, Department of Physics, Guwahati, India
\item \Idef{org43}Helmholtz-Institut f\"{u}r Strahlen- und Kernphysik, Rheinische Friedrich-Wilhelms-Universit\"{a}t Bonn, Bonn, Germany
\item \Idef{org44}Helsinki Institute of Physics (HIP), Helsinki, Finland
\item \Idef{org45}Hiroshima University, Hiroshima, Japan
\item \Idef{org46}Hochschule Worms, Zentrum  f\"{u}r Technologietransfer und Telekommunikation (ZTT), Worms, Germany
\item \Idef{org47}Horia Hulubei National Institute of Physics and Nuclear Engineering, Bucharest, Romania
\item \Idef{org48}Indian Institute of Technology Bombay (IIT), Mumbai, India
\item \Idef{org49}Indian Institute of Technology Indore, Indore, India
\item \Idef{org50}Indonesian Institute of Sciences, Jakarta, Indonesia
\item \Idef{org51}INFN, Laboratori Nazionali di Frascati, Frascati, Italy
\item \Idef{org52}INFN, Sezione di Bari, Bari, Italy
\item \Idef{org53}INFN, Sezione di Bologna, Bologna, Italy
\item \Idef{org54}INFN, Sezione di Cagliari, Cagliari, Italy
\item \Idef{org55}INFN, Sezione di Catania, Catania, Italy
\item \Idef{org56}INFN, Sezione di Padova, Padova, Italy
\item \Idef{org57}INFN, Sezione di Roma, Rome, Italy
\item \Idef{org58}INFN, Sezione di Torino, Turin, Italy
\item \Idef{org59}INFN, Sezione di Trieste, Trieste, Italy
\item \Idef{org60}Inha University, Incheon, Republic of Korea
\item \Idef{org61}Institut de Physique Nucl\'{e}aire d'Orsay (IPNO), Institut National de Physique Nucl\'{e}aire et de Physique des Particules (IN2P3/CNRS), Universit\'{e} de Paris-Sud, Universit\'{e} Paris-Saclay, Orsay, France
\item \Idef{org62}Institute for Nuclear Research, Academy of Sciences, Moscow, Russia
\item \Idef{org63}Institute for Subatomic Physics, Utrecht University/Nikhef, Utrecht, Netherlands
\item \Idef{org64}Institute for Theoretical and Experimental Physics, Moscow, Russia
\item \Idef{org65}Institute of Experimental Physics, Slovak Academy of Sciences, Ko\v{s}ice, Slovakia
\item \Idef{org66}Institute of Physics, Bhubaneswar, India
\item \Idef{org67}Institute of Physics of the Czech Academy of Sciences, Prague, Czech Republic
\item \Idef{org68}Institute of Space Science (ISS), Bucharest, Romania
\item \Idef{org69}Institut f\"{u}r Kernphysik, Johann Wolfgang Goethe-Universit\"{a}t Frankfurt, Frankfurt, Germany
\item \Idef{org70}Instituto de Ciencias Nucleares, Universidad Nacional Aut\'{o}noma de M\'{e}xico, Mexico City, Mexico
\item \Idef{org71}Instituto de F\'{i}sica, Universidade Federal do Rio Grande do Sul (UFRGS), Porto Alegre, Brazil
\item \Idef{org72}Instituto de F\'{\i}sica, Universidad Nacional Aut\'{o}noma de M\'{e}xico, Mexico City, Mexico
\item \Idef{org73}iThemba LABS, National Research Foundation, Somerset West, South Africa
\item \Idef{org74}Johann-Wolfgang-Goethe Universit\"{a}t Frankfurt Institut f\"{u}r Informatik, Fachbereich Informatik und Mathematik, Frankfurt, Germany
\item \Idef{org75}Joint Institute for Nuclear Research (JINR), Dubna, Russia
\item \Idef{org76}Korea Institute of Science and Technology Information, Daejeon, Republic of Korea
\item \Idef{org77}KTO Karatay University, Konya, Turkey
\item \Idef{org78}Laboratoire de Physique Subatomique et de Cosmologie, Universit\'{e} Grenoble-Alpes, CNRS-IN2P3, Grenoble, France
\item \Idef{org79}Lawrence Berkeley National Laboratory, Berkeley, California, United States
\item \Idef{org80}Lund University Department of Physics, Division of Particle Physics, Lund, Sweden
\item \Idef{org81}Nagasaki Institute of Applied Science, Nagasaki, Japan
\item \Idef{org82}Nara Women{'}s University (NWU), Nara, Japan
\item \Idef{org83}National and Kapodistrian University of Athens, School of Science, Department of Physics , Athens, Greece
\item \Idef{org84}National Centre for Nuclear Research, Warsaw, Poland
\item \Idef{org85}National Institute of Science Education and Research, HBNI, Jatni, India
\item \Idef{org86}National Nuclear Research Center, Baku, Azerbaijan
\item \Idef{org87}National Research Centre Kurchatov Institute, Moscow, Russia
\item \Idef{org88}Niels Bohr Institute, University of Copenhagen, Copenhagen, Denmark
\item \Idef{org89}Nikhef, National institute for subatomic physics, Amsterdam, Netherlands
\item \Idef{org90}NRC ¿Kurchatov Institute¿ ¿ IHEP , Protvino, Russia
\item \Idef{org91}NRNU Moscow Engineering Physics Institute, Moscow, Russia
\item \Idef{org92}Nuclear Physics Group, STFC Daresbury Laboratory, Daresbury, United Kingdom
\item \Idef{org93}Nuclear Physics Institute of the Czech Academy of Sciences, \v{R}e\v{z} u Prahy, Czech Republic
\item \Idef{org94}Oak Ridge National Laboratory, Oak Ridge, Tennessee, United States
\item \Idef{org95}Petersburg Nuclear Physics Institute, Gatchina, Russia
\item \Idef{org96}Physics department, Faculty of science, University of Zagreb, Zagreb, Croatia
\item \Idef{org97}Physics Department, Panjab University, Chandigarh, India
\item \Idef{org98}Physics Department, University of Jammu, Jammu, India
\item \Idef{org99}Physics Department, University of Rajasthan, Jaipur, India
\item \Idef{org100}Physikalisches Institut, Eberhard-Karls-Universit\"{a}t T\"{u}bingen, T\"{u}bingen, Germany
\item \Idef{org101}Physikalisches Institut, Ruprecht-Karls-Universit\"{a}t Heidelberg, Heidelberg, Germany
\item \Idef{org102}Physik Department, Technische Universit\"{a}t M\"{u}nchen, Munich, Germany
\item \Idef{org103}Research Division and ExtreMe Matter Institute EMMI, GSI Helmholtzzentrum f\"ur Schwerionenforschung GmbH, Darmstadt, Germany
\item \Idef{org104}Rudjer Bo\v{s}kovi\'{c} Institute, Zagreb, Croatia
\item \Idef{org105}Russian Federal Nuclear Center (VNIIEF), Sarov, Russia
\item \Idef{org106}Saha Institute of Nuclear Physics, Kolkata, India
\item \Idef{org107}School of Physics and Astronomy, University of Birmingham, Birmingham, United Kingdom
\item \Idef{org108}Secci\'{o}n F\'{\i}sica, Departamento de Ciencias, Pontificia Universidad Cat\'{o}lica del Per\'{u}, Lima, Peru
\item \Idef{org109}Shanghai Institute of Applied Physics, Shanghai, China
\item \Idef{org110}Stefan Meyer Institut f\"{u}r Subatomare Physik (SMI), Vienna, Austria
\item \Idef{org111}SUBATECH, IMT Atlantique, Universit\'{e} de Nantes, CNRS-IN2P3, Nantes, France
\item \Idef{org112}Suranaree University of Technology, Nakhon Ratchasima, Thailand
\item \Idef{org113}Technical University of Ko\v{s}ice, Ko\v{s}ice, Slovakia
\item \Idef{org114}Technische Universit\"{a}t M\"{u}nchen, Excellence Cluster 'Universe', Munich, Germany
\item \Idef{org115}The Henryk Niewodniczanski Institute of Nuclear Physics, Polish Academy of Sciences, Cracow, Poland
\item \Idef{org116}The University of Texas at Austin, Austin, Texas, United States
\item \Idef{org117}Universidad Aut\'{o}noma de Sinaloa, Culiac\'{a}n, Mexico
\item \Idef{org118}Universidade de S\~{a}o Paulo (USP), S\~{a}o Paulo, Brazil
\item \Idef{org119}Universidade Estadual de Campinas (UNICAMP), Campinas, Brazil
\item \Idef{org120}Universidade Federal do ABC, Santo Andre, Brazil
\item \Idef{org121}University College of Southeast Norway, Tonsberg, Norway
\item \Idef{org122}University of Cape Town, Cape Town, South Africa
\item \Idef{org123}University of Houston, Houston, Texas, United States
\item \Idef{org124}University of Jyv\"{a}skyl\"{a}, Jyv\"{a}skyl\"{a}, Finland
\item \Idef{org125}University of Liverpool, Department of Physics Oliver Lodge Laboratory , Liverpool, United Kingdom
\item \Idef{org126}University of Split, Faculty of Electrical Engineering, Mechanical Engineering and Naval Architecture, Split, Croatia
\item \Idef{org127}University of Tennessee, Knoxville, Tennessee, United States
\item \Idef{org128}University of the Witwatersrand, Johannesburg, South Africa
\item \Idef{org129}University of Tokyo, Tokyo, Japan
\item \Idef{org130}University of Tsukuba, Tsukuba, Japan
\item \Idef{org131}Universit\'{e} Clermont Auvergne, CNRS/IN2P3, LPC, Clermont-Ferrand, France
\item \Idef{org132}Universit\'{e} de Lyon, Universit\'{e} Lyon 1, CNRS/IN2P3, IPN-Lyon, Villeurbanne, Lyon, France
\item \Idef{org133}Universit\'{e} de Strasbourg, CNRS, IPHC UMR 7178, F-67000 Strasbourg, France, Strasbourg, France
\item \Idef{org134} Universit\'{e} Paris-Saclay Centre d¿\'Etudes de Saclay (CEA), IRFU, Department de Physique Nucl\'{e}aire (DPhN), Saclay, France
\item \Idef{org135}Universit\`{a} degli Studi di Pavia, Pavia, Italy
\item \Idef{org136}Universit\`{a} di Brescia, Brescia, Italy
\item \Idef{org137}V.~Fock Institute for Physics, St. Petersburg State University, St. Petersburg, Russia
\item \Idef{org138}Variable Energy Cyclotron Centre, Kolkata, India
\item \Idef{org139}Warsaw University of Technology, Warsaw, Poland
\item \Idef{org140}Wayne State University, Detroit, Michigan, United States
\item \Idef{org141}Westf\"{a}lische Wilhelms-Universit\"{a}t M\"{u}nster, Institut f\"{u}r Kernphysik, M\"{u}nster, Germany
\item \Idef{org142}Wigner Research Centre for Physics, Hungarian Academy of Sciences, Budapest, Hungary
\item \Idef{org143}Yale University, New Haven, Connecticut, United States
\item \Idef{org144}Yonsei University, Seoul, Republic of Korea
\end{Authlist}
\endgroup

%
\end{document}